\documentclass[usenatbib]{mn2e}
\usepackage{graphicx}

\newcommand{\kms}{$\;$km s$^{-1}$}
\newcommand{\msun}{M_{\odot}}

\begin{document}
 
\title[A high relative-precision CMD of M67]{A high relative precision
color-magnitude diagram of M67}

\author[E. L. Sandquist]{Eric L. Sandquist$^{1}$\thanks{E-mail:
erics@mintaka.sdsu.edu}\\ $^{1}$San Diego State University,Department
of Astronomy,San Diego, CA 92182 USA}

\date{}

\maketitle

\begin{abstract}
We have calibrated and combined an extensive set of $BVI$ observations
of M67 to produce a color-magnitude diagram of stars measured with
high relative precision. We have selected stars that are most likely
to be single-star members of the cluster using proper motion, radial
velocity, and variability information from the literature, and an
examination of the most probable color-magnitude diagram locations of
unresolved stellar blends. We have made detailed comparisons of our
photometry of the selected stars with theoretical models, and discuss
the most notable discrepancies. Observations of M67 turnoff stars are
a severe test of algorithms attempting to describe convective cores in
the limit of small extent, and we find strong evidence of a ``hook''
just fainter than the turnoff gap.  The stars in M67 support
assertions that the degree of convective core overshooting decreases
to zero for stars with masses in the range $1.0 < (M / \msun) \le
1.5$, but that the degree of overshoot is smaller than currently used
in published isochrones. We also verify that all current theoretical
models for the lower main sequence (with the exception of Baraffe et
al. 1998) are too blue for $M_V \ga 6$ even when the sequences are
shifted to match M67 near the $M_V$ of the Sun, probably due to a
combination of problems with color-$T_{\mbox{eff}}$ transformations
and realistic surface boundary conditions for models. Finally, we
identify a subset of cluster members with unusual photometry
(candidate red giant binaries, blue straggler stars, and triple
systems) deserving of further study.
\end{abstract}


\section{Introduction}
 
Color-magnitude diagrams (CMDs) of stellar clusters have long been
used as tests of stellar evolution because of the way they constrain
the evolution of the luminosity and surface temperature of individual
stars. A number of difficulties have to be overcome to ensure the
best quality comparisons with theoretical models, and one of these is
obtaining a sufficient number of measurements of stellar fluxes in
different filter bands. With small numbers of observations,
measurement scatter is often the most important issue since the
position of a star in the CMD will not simply be dependent on its
mass, chemical composition, and evolutionary state. In a cluster of
stars, the ensemble of measured stars must then be used to infer where
a single star would fall in the diagram. However, with a large enough
number of independent observations of a group of stars, the
photometric scatter can be reduced to the point where it is no longer
a significant contributor to uncertainty in the position of the
cluster's fiducial line. At that point, scatter introduced by
unresolved binaries, variable stars, and nonmembers become the main
contributors to the scatter.

With a high-precision photometric database, calibration to a standard
system can require relatively little additional effort: a few nights
of observations under photometric conditions with fairly large numbers
of cluster star observations interspersed with standard star
observations covering the color range of the cluster stars to be
calibrated and covering the airmass range of the cluster
observations. Of course, the standard calibration can also be improved
with additional data as long as the standard system is well determined.

There are numerous benefits to high precision CMD studies. First, the
shapes of the fiducial lines often reveal inadequacies in our
understanding of the stellar physics involved in modeling the
stars. An example is the turnoff region of open clusters, where
observations have shown that a small amount of convective overshooting
in the convective cores of stars is necessary to explain the
morphology of the fiducial line. High precision measurements can also
reveal the presence of peculiar stars. Two examples are the poorly
understood ``sub-subgiant branch'' stars S1063 and S1113 in M67
(Mathieu et al. 2003; identifications starting with ``S'' are from
Sanders 1977). Both stars are high-probability cluster members, but
fall in a region of the CMD where it is very difficult to explain
their photometric properties.

High-precision photometry can also open up the possibility of
spectroscopic studies of the surface properties and abundances by
identifying samples of probable single stars. While proper motion
studies can identify cluster members, eliminating unresolved binaries
from the sample generally requires large investments of time for
radial velocity studies. Even so, wide binaries and binaries with
extreme mass ratios are unlikely to be identified except with the help
of precise photometry. Field star contamination of cluster CMDs can
often make the selection of cluster members difficult if proper motion
studies have not been carried out. Even when they have been, the first
epoch observations were most frequently taken using photographic
techniques, which most often means that faint main sequence stars are
unlikely to have been included.

These issues are particularly pressing in the case of open clusters
like M67.  M67 is probably the most thoroughly studied old open
cluster in the Galaxy, thanks to its small distance from us. Typically
quoted values for the cluster's age ($4 \pm 0.5$ Gyr; Dinescu et
al. 1995) place it between the majority of known open clusters and the
much older globular clusters. There have also been a number of proper
motion membership studies of the cluster \citep{sanders,girard,zhao}
and radial velocity studies \citep{mathieurv,mathieuorb} that can be
used to help ``clean'' the CMD of nonmembers and binary stars in order
to make the single-star sequence more apparent.  A number of authors
have commented on the seemingly large binary star fraction in the
cluster \citep{mmj,fan}, which probably indicates a high degree of
mass segregation and may also indicate that the cluster is in the late
stages of its dynamical life during which it is being tidally
disrupted.

Recent attention on clusters has come from variability studies, and in
some cases a by-product of such a study can be a large number of
observations of non-varying stars. The dataset we describe below is a
result of studies of two particularly difficult eclipsing variables in
M67: the blue straggler S1082 ($P =$ 1.0677978 d for the eclipsing
binary), and the totally-eclipsing variable S986 ($P =$ 10.33813 d).
We gathered thousands of frames of observations of M67 in $V$ band,
hundreds of observations in $I$, and tens in $B$.  We will focus our
attention on the $VI$ CMD because of its high precision.  \citet{fan}
has also presented wide-field photometry of M67 in nine filters. Their
photometry in at least three of those filters (with bandpasses
centered at 3890, 6075, and 9190 \AA) certainly qualifies as high
precision. However, because their photometry was taken in a
non-standard filter system, the comparison of their data with
theoretical values is more complicated than ours. However, we will
use their data in conjunction with ours to attempt to identify the
stars that have the highest probability of being single stars.

We briefly describe our photometry and its reduction and calibration
in \S 2. In \S 3, we present our identification of the single star
sequence and determination of the fiducial line. In \S 4, we compare
the fiducial line with theoretical isochrones.

\section{Photometric Observations and Reductions}

All of the photometry for this study was taken at the 1 m telescope at
the Mt. Laguna Observatory using a $2048 \times 2048$ CCD on nights
between December 2000 and April 2003. The nights of observations
are given in Table \ref{obs}. Typical exposure times ranged between 10
and 60 s to optimize the counts for our eclipsing binary targets near
the cluster turnoff.

\begin{table}
\caption{Observing Log for Photometry at Mt. Laguna}
\label{obs}
\begin{tabular}{@{}clccl}
\hline
\# & Date & Filters & mJD Start & $N_{obs}$ \\
\hline
1 & Jan. 27/28, 2000 & $BVI$ & 1571.739 & 22,20,23 \\
2 & Dec. 5/6, 2000 & $V$ & 1884.878 & 60\\
3 & Dec. 6/7, 2000 & $I$ & 1885.871 & 38\\
4 & Dec. 7/8, 2000 & $V$ & 1886.806 & 111\\
5 & Dec. 11/12, 2000 & $V$ & 1890.817 & 57\\
6 & Dec. 12/13, 2000 & $V$ & 1891.820 & 116\\
7 & Dec. 13/14, 2000 & $B$ & 1892.815 & 28\\
8 & Jan. 23/24, 2001 & $V$ & 1933.675 & 206\\
9 & Jan. 25/26, 2001 & $V$ & 1935.673 & 176\\
10 & Jan. 29/30, 2001 & $V$ & 1939.676 & 44\\
11 & Jan. 30/31, 2001 & $V$ & 1940.657 & 130\\
12 & Jan. 31/Feb. 1, 2001 & $V$ & 1941.663 & 161\\
13 & Feb. 17/18, 2001 & $V$ & 1958.655 & 63\\
14 & Feb. 18/19, 2001 & $V$ & 1959.790 & 75\\
15 & Nov. 14/15, 2001 & $V$ & 2228.901 & 74\\
16 & Jan. 20/21, 2002 & $V$ & 2295.735 & 91\\
17 & Jan. 21/22, 2002 & $V$ & 2296.687 & 89\\
18 & Jan. 24/25, 2002 & $V$ & 2299.690 & 124\\
19 & Feb. 5/6, 2002 & $V$ & 2311.649 & 119\\
20 & Feb. 10/11, 2002 & $V$ & 2316.638 & 167\\
21 & Mar. 18/19, 2002 & $BV$ & 2352.625 & 12,131\\
22 & Apr. 13/14, 2002 & $V$ & 2378.638 & 103\\
23 & Apr. 18/19, 2002 & $V$ & 2383.661 & 47\\
24 & Nov. 21/22, 2002 & $VI$ & 2600.845 & 108,25\\
25 & Jan. 17/18, 2003 & $I$ & 2657.718 & 192\\
26 & Jan. 22/23, 2003 & $I$ & 2662.702 & 110\\
27 & Feb. 17/18, 2003 & $I$ & 2688.617 & 147\\
28 & Mar. 20/21, 2003 & $I$ & 2719.635 & 138\\
29 & Apr. 20/21, 2003 & $I$ & 2750.638 & 150\\
\hline
\end{tabular}

\medskip
Note: mJD = HJD - 2450000
\end{table}

Most of the details of the reduction are presented in other papers
(Sandquist et al. 2003a, Sandquist \& Shetrone 2003b), so we only
briefly describe the reduction here.  The object frames were reduced
in usual fashion, using overscan subtraction, bias frames, and flat
fields.  We conducted aperture photometry using the IRAF\footnote{IRAF
(Image Reduction and Analysis Facility) is distributed by the National
Optical Astronomy Observatories, which are operated by the Association
of Universities for Research in Astronomy, Inc., under contract with
the National Science Foundation.} tasks DAOFIND and PHOT from the
APPHOT package.  In order to improve the accuracy of the relative
photometry for the light curves, we used an ensemble photometry method
similar to that described by Honeycutt (1992), iterating toward a
consistent solution for photometric zeropoints for all frames and
median magnitudes for all stars. One major difference with other
ensemble photometry techniques involved the use of position-dependent
corrections to stellar magnitudes to account for variations in the
point-spread function across the frame and for changes in frame
centre. Because frames typically had more than 300 measurable stars,
the formal errors in the zero points ranged from around 0.003 to 0.007
mag, even with respect to night-to-night variations. Because of the
often large number of observations of individual stars, typical errors
in star magnitudes were a few millimag to a fraction of a
millimag. (Because we relied on robust median magnitudes, these errors
were derived in the following way. We first created a list of
individual observations of a given star in a given bandpass and found
the median. We then took the square-root of the number of observations
of the star in the bandpass, and looked up the observations of the
star that number of entries away from the median entry in the list
ordered by magnitude. The error was taken to be one half of the
difference between those observations.)

\subsection{Photometric Calibration}

We have completed a calibration of our photometric dataset to attempt
to ensure accurate placement of our photometry on the standard
system. The large numbers of measurements in our studies of M67 has
meant that the main sequence scatter is smaller than in either
\citet{mmj} or the more recent Richer et al. (1998), which both used
standard Johnson-Cousins filters. Our calibration was taken from a
night of photometric data taken at Mount Laguna Observatory on January
27/28, 2000. Observations of the Landolt standard fields SA 95, 97,
98, 101, and 107 were interspersed with observations of M67
through the $B$, $V$, and $I$ filters (at least 20 images in each
band). We used Stetson (2000) standard values for stars when
available, and Landolt (1992) values for the remainder of the
standards.

Aperture photometry was performed on both standard and cluster frames
using DAOPHOT II (Stetson 1987) using multiple synthetic
apertures. Growth curves were used to extrapolate measurements to a
(large) common aperture size using the program DAOGROW (Stetson 1990).
The photometric transformation equations used in the calibration were
\[ b = B + a_0 + a_1 (B-V) + a_2 (X - 1.25)\]
\[ v = V + b_0 + b_1 (V-I) + b_2 (X - 1.25)\]
\[ i = I + c_0 + c_1 (V-I) + c_2 (X - 1.25) + c_3 (V-I)^2 \] \[+ c_4 (V-I)^3\]
where $b$, $v$, and $i$ are the observed aperture photometry
magnitudes, $B$, $V$, and $I$ are the standard system magnitudes, and
$X$ is airmass. The transformation coefficients were determined using
the program CCDSTD (e.g. Stetson 1992), and with the exception of the
$I$-band data, the data were well fitted by linear color terms.

\begin{figure}
\includegraphics[width=84mm]{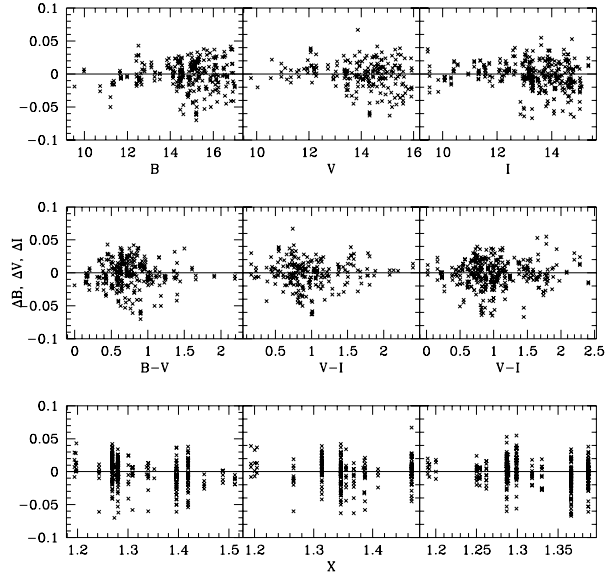}
\caption{Residuals (in the sense of our observed values minus the standard
values) from the calibration of primary photometric standards versus
magnitude (top row), color (middle row), and airmass $X$ (bottom
row). The first column is for $B$ measurements, the second for $V$,
and the third for $I$.
\label{primary}}
\end{figure}

\begin{figure}
\includegraphics[width=84mm]{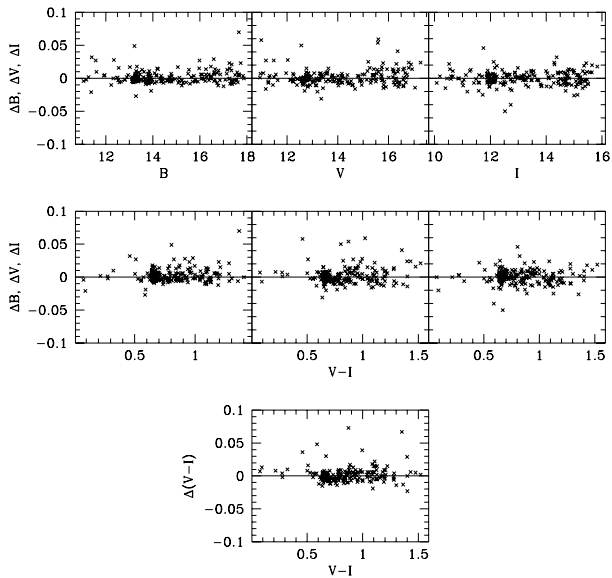}
\caption{Residuals (in the sense of our observed values minus the standard
values) from the calibration of secondary photometric standards in M67
versus magnitude (top row), color (middle row), and color residuals
versus color (bottom row).
\label{secondary}}
\end{figure}

We selected 197 stars in the central part of the cluster to be our
secondary standards. These stars were generally on the lower RGB,
subgiant branch, upper MS, or blue stragglers, and covered most of the
range of colors for the cluster stars observed ($0 \la (V-I) \la
1.5$). We noticed that the secondary standard observations, once
corrected to the standard system, had significant trends with
airmass. Since the M67 stars were observed over a wider range of
airmasses than the primary standards, we used these trends to derive
corrected airmass terms for the transformation equations. The
corrected values removed the airmass trends, and did not noticeably
affect the quality of the fits to the primary standards, so we adopted
the corrected airmass terms for the transformations.  The residual
plots for the primary standard fits are shown in Fig. \ref{primary},
and the transformation coefficients are given in Table \ref{trans}.
The calibrated secondary standard values were then used to calibrate
our ensemble photometry. We present the transformation coefficients in
Table \ref{trans}. Because physically different filters were used on
the night of calibration and during the ensemble photometry, there are
significant color terms and there is a significant second-order term
present in the $I$-band residuals. The results of the calibration are
shown in Fig. \ref{secondary}. We must emphasize that we can only be
confident of the calibration within the range of colors of our
secondary standards, and that this is {\it not} the entire range of
colors covered by our ensemble photometry, so that there may be
systematic errors in the absolute calibration of the lower main
sequence. For the purposes of the distance modulus determination, our
photometry appears to be accurately calibrated to the standard system.

\begin{table*}
\centering
\begin{minipage}{14cm}
\caption{Photometric Transformation Equation Coefficients}
\label{trans}
\begin{tabular}{@{}ccccc@{}}
\hline
Filter & \multicolumn{2}{c}{Airmass Term} & \multicolumn{2}{c}{Color Term} \\
\hline
\multicolumn{5}{c}{\underline{Primary Calibration}} \\
$B$ & Uncorrected & $0.1326\pm0.0120$ & First Order & $-0.0708\pm0.0027$ \\
  & Corrected & $0.2168\pm0.0121$ & Corrected & $-0.0683\pm0.0030$ \\
$V$ & Uncorrected & $0.0808\pm0.0154$ & First Order & $0.0067\pm0.0027$ \\
 & Corrected & $0.1103\pm0.0156$ & Corrected & $0.0072\pm0.0026$ \\
$I$ & Uncorrected & 0. & First Order & $0.0623\pm0.0150$ \\
 & & & Second Order & $-0.0888\pm0.0148$ \\
 & & & Third Order & $0.0214\pm0.0043$\\
 & Corrected & $0.0505\pm0.0013$ & First Order & $0.0779\pm0.0156$\\
 & & & Second Order & $-0.1020\pm0.0153$ \\
 & & & Third Order & $0.0244\pm0.0045$\\
\multicolumn{5}{c}{\underline{Secondary Calibration}} \\
$B$ & & & First Order & $-0.0615\pm0.0024$\\
$V$ & & & First Order & $-0.0346\pm0.0024$\\
$I$ & & & First Order & $0.0332\pm0.0113$\\
 & & & Second Order & $-0.0373\pm0.0065$\\
\hline
\end{tabular}
\end{minipage}
\end{table*}

\begin{figure}
\includegraphics[width=84mm]{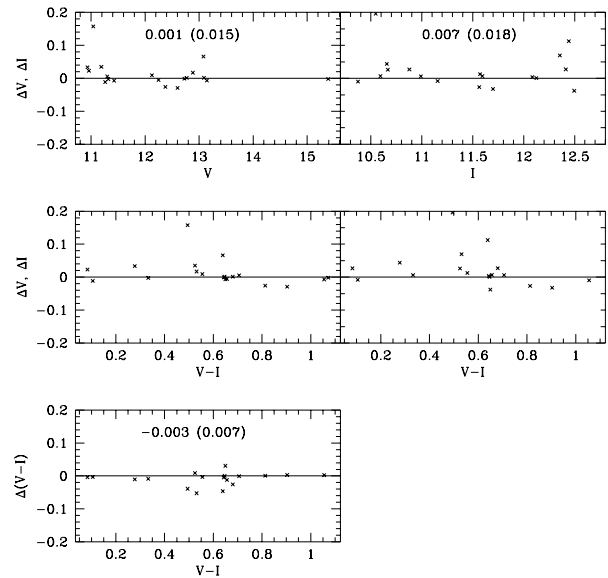}
\caption{Residuals (in the sense of our observed values minus the
published values) from the comparison of calibrated M67 photometry
with the study of \citet{jt}. Also
included are the median residual values and in parentheses the
semi-interquartile range (a measure of dispersion).
\label{jtfig}}
\end{figure}

\begin{figure}
\includegraphics[width=84mm]{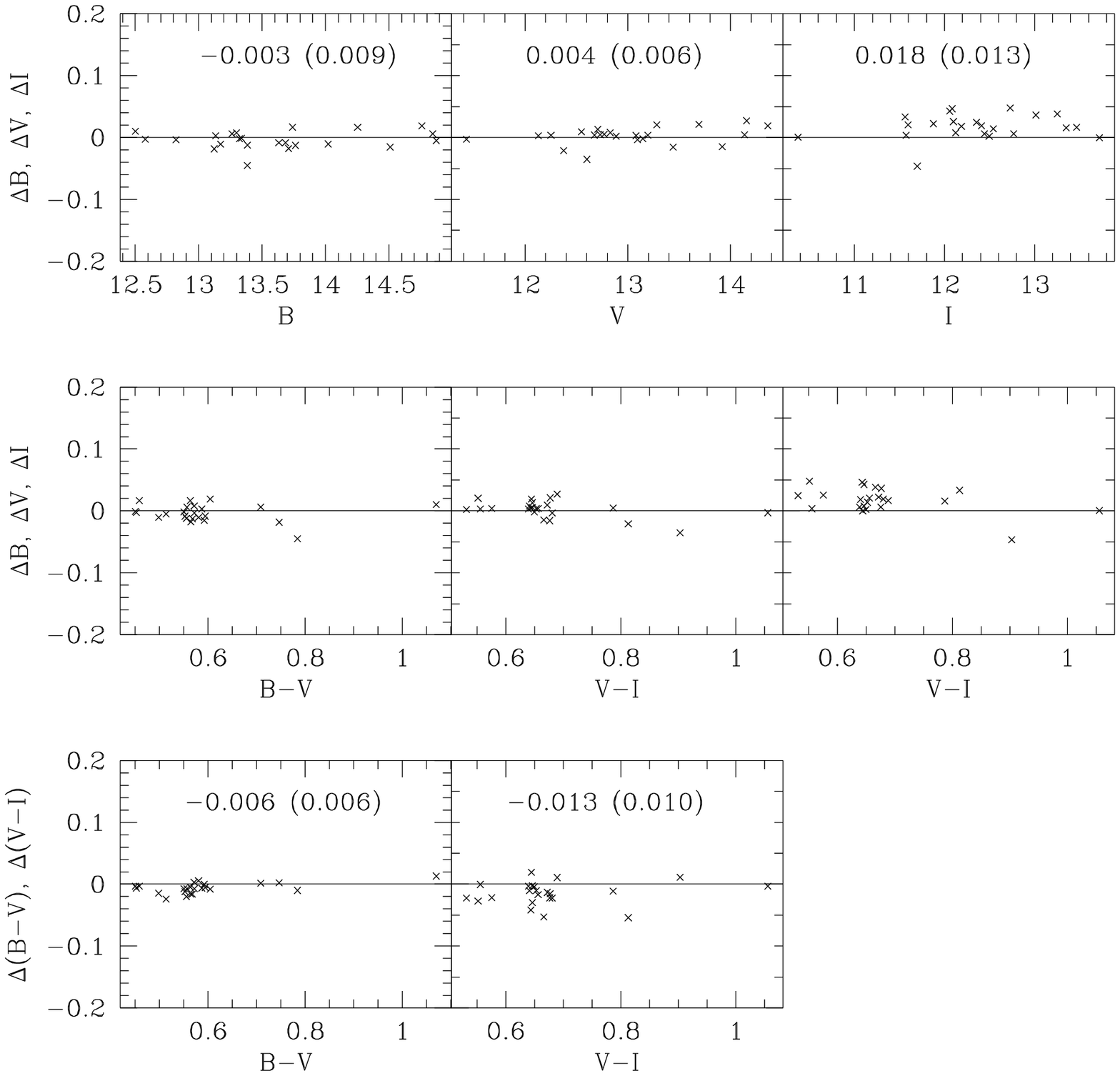}
\caption{Residuals (in the sense of our observed values minus the
published values) from the comparison of calibrated M67 photometry
with the study of \citet{ci}.
\label{cifig}}
\end{figure}

We made comparisons between our dataset and those of \citet{jt} and
\citet{ci} (both studies made detailed attempts to tie their
photometry to the Landolt standard system), and to \citet{mmj}, which
is the most frequently cited source of M67 photometry. We find good
agreement between our photometry and values from \citet{jt} and
\cite{ci} as shown in Figs. \ref{jtfig} and \ref{cifig}. One exception
is the slight zero-point difference in $I$ between our photometry and
that of \citet{ci}. We find more substantial differences between our
photometry and that of \citet{mmj}, however. In the color range
covered by the photometry of \citet{jt} (which Montgomery et al. used
in calibrating their photometry), the agreement is fairly good
(although there are significant zero-point offsets). However, outside
of that range, the color-dependent residuals become noticeable, and
there is a clear color-dependent trend in the $(V-I)$ residuals. Because
both our primary and secondary standards covered colors up to $(V-I)
\approx 1.5$, we believe our photometry should be more reliable.

\begin{figure}
\includegraphics[width=84mm]{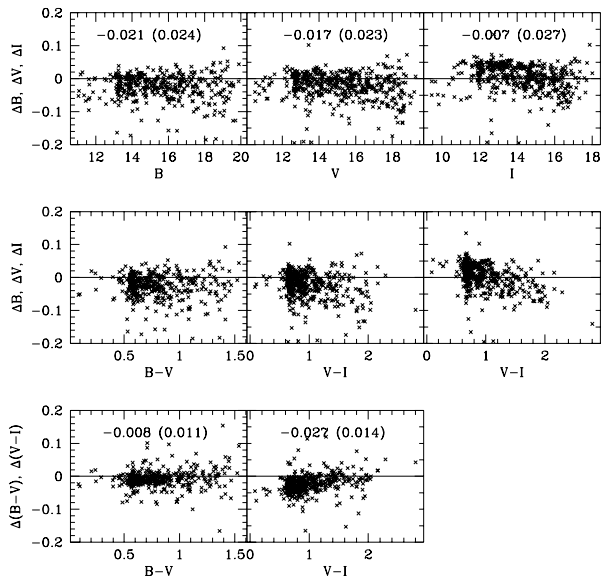}
\caption{Residuals (in the sense of our observed values minus the
published values) from the comparison of calibrated M67 photometry
with the study of \citet{mmj}.
\label{mmjfig}}
\end{figure}

\section{The Single-Star Sequence and Fiducial Line}

For the best possible comparison with theoretical isochrones, we would
ideally like to identify a sample of unambiguous single-star cluster
members that have been able to evolve without having undergone strong
interactions with their neighbors. We will refer to the locus of the
positions of these stars in the color-magnitude diagram (CMD) as the
single-star sequence (SSS). To define such a sequence, contaminants in
the CMD have to be removed. For a relatively sparse cluster like M67,
this task is particularly important, so we describe our selection
criteria in detail below.

\subsection{Selection Criteria}

\begin{figure*}
\includegraphics[width=160mm]{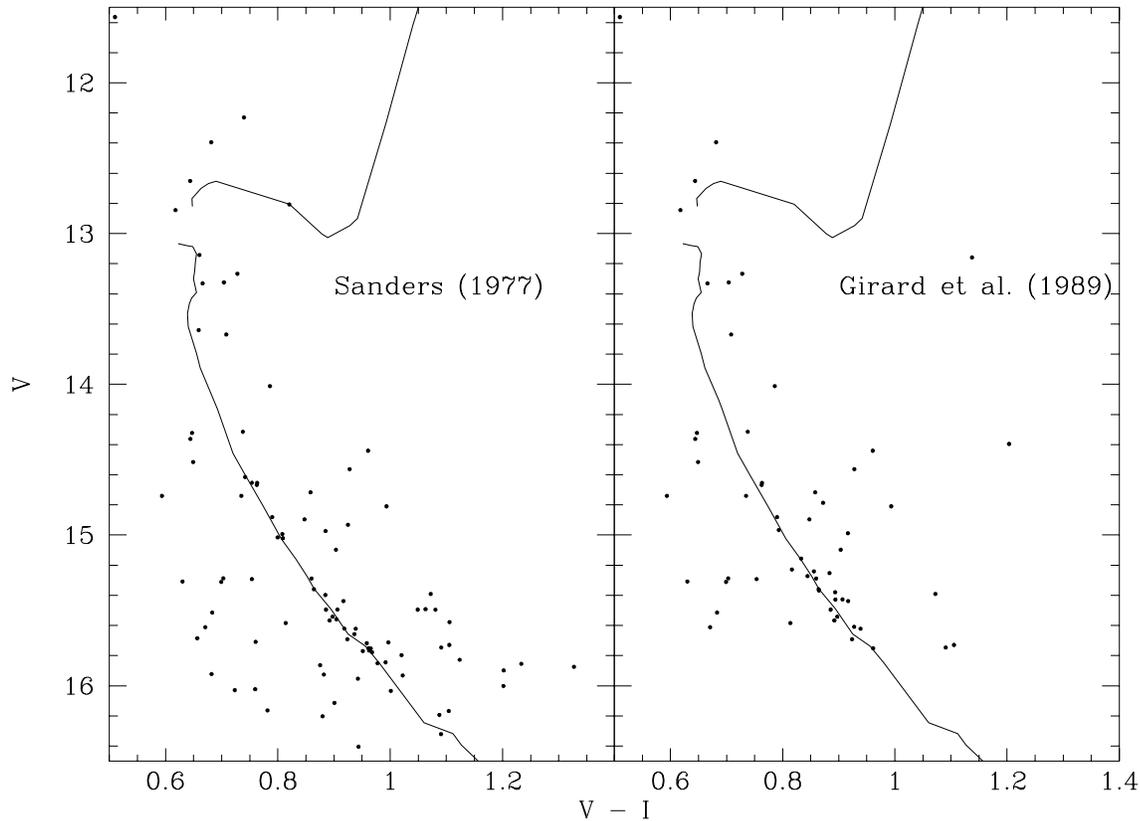}
\caption{CMD of cluster non-members (membership probability $< 50$\%)
according to proper motion studies. The
photometry is from this study.
\label{nonmem}}
\end{figure*}

The primary source of contamination in the CMD on the lower main
sequence is from field stars. We have used proper motion membership
probabilities from \citet{sanders} and \citet{girard} for guidance
where possible. Both studies had a faint limit of $V \approx 16$.
although we believe that the membership probabilities for the fainter
stars in the Sanders sample may be systematically underestimated, as
we found that an abnormally large number of stars with excellent $VI$
data clearly falling on the main sequence (as delineated by other
higher probability members) had low membership probabilities (see
Fig. \ref{nonmem}). In particular, there are a large number of
low-probability members that fall near the main sequence having $15.2
< V < 16$. This seems to be the case for the Girard et al. dataset to a
similar degree, so we will not rely entirely on the proper motions for
faint stars except when both studies agree that a star is a member.

Due to the high binary content of M67 \citep{mmj,fan}, attention has
to be paid to contamination of the CMD by multiple star systems. We
conducted a literature search for confirmed star systems among the
cluster members, and found that the elimination of these systems from
the CMD clarifies the position of the SSS near the cluster turnoff
(TO), subgiant branch (SGB), and lower red giant branch (RGB). Known
multiple star systems are identified in Table \ref{binary} along with
the most recent orbit information available and photometry from this
study, \citet{mmj} (for S1040, S1113, S1195, and S1250), or
\citet{SvMV} (for S1508).

\begin{table}
\centering
\caption{Known Multiple Star Systems in M67}
\label{binary}
\begin{tabular}{@{}lrclcc@{}}
\hline
ID & $V$ & $V-I$ & $P$ (d) & $e$ & Ref.\\
\hline
EU Cnc & & & 0.09 & 0. & 6 \\
III-79& 15.87 & 1.26 & 0.2704 & 0 & 6 \\
S757  & 13.49 & 0.67 & 0.35967 & 0 & 5 \\
S1282 & 13.44 & 0.67 & 0.360452 & 0 & 7 \\
S1036 & 12.80 & 0.59 & 0.441437 & 0 & 5 \\
S1082 & 11.19 & 0.53 & 1.067797 & 0 & 2 \\
S972  & 15.39 & 1.07 & 1.166412 & 0.01 & 10 \\
S1077 & 12.61 & 0.81 & 1.358766 & 0.10 & 7 \\
S1019 & 14.26 & 0.89 & 1.36022 & 0.02 & 7 \\
S1070 & 13.93 & 0.73 & 2.66 & 0 & 10 \\
S1113 & 13.77 & & 2.823094 & 0 & 8 \\
S1234 & 12.62 & 0.66 & 4.18258 & 0.26 & 1\\
S1024 & 12.71 & 0.65 & 7.15961 & 0 & 4 \\
S1045 & 12.56 & 0.67 & 7.6452 & 0 & 4 \\
S999  & 12.60 & 0.90 & 10.0553 & 0 & 4 \\
S986  & 12.73 & 0.64 & 10.33813 & 0 & 3 \\
S1272 & 12.54 & 0.66 & 11.0215 & 0.26 & 4 \\
S2206 & 12.38 & 0.81 & 18.377 & 0.34 & 4 \\
S1063 & 13.52 & 0.65 & 18.396 & 0.21 & 8 \\
S1508 & 13.45 & 1.31 & 25.866 & 0.44 & 4 \\
S821  & 12.79 & 0.68 & 26.259 & 0.4 & 9 \\
S1242 & 12.67 & 0.76 & 31.780 & 0.66 & 4 \\
S1040 & 11.52 & 0.91 & 42.877 & 0. & 4 \\
S1216 & 12.67 & 0.65 & 60.445 & 0.45 & 4 \\
S1053 & 12.23 & 0.74 & 123.39 & 0.49 & 4, mem?\\
S1285 & 12.51 & 0.73 & 277.8 & 0.19 & 4 \\
S1264a & 12.05 & 1.03 & 353.9 & 0.38 & 4 \\
S1000 & 12.81 & 0.82 & 531 & 0.09 & 4, nm \\
S1237 & 10.73 & 0.96 & 697.8 & 0.11 & 4 \\
S1267 & 10.91 & 0.24 & 850 & 0.47 & 1 \\
S251  & & & 948 & 0.55 & 4 \\
S760  & 13.29 & 0.68 & 954 & 0.43 & 7 \\
S752  & 11.32 & 0.33 & 1013 & 0.27 & 1 \\
S1195 & 12.28 & 0.50 & 1139 & 0.03 & 1 \\
S975  & 11.04 & 0.49 & 1231 & 0.12 & 1 \\
S1182 & & & 1233 & 0.13 & 4 \\
S440  & & & 1315 & 0.15 & 4 \\
S1072 & 11.30 & 0.71 & 1495 & 0.32 & 4\\
S1250 &  9.69 & 1.33 & 4410 & 0.50 & 4 \\
S997  & 12.13 & 0.56 & 5153 & 0.36 & 1 \\
S1221 & 10.74 & 1.09 & 6445 & 0 & 4 \\
\hline
\end{tabular}

\medskip
Notes: mem?: questionable cluster member; nm: low membership probability

References: 1: Sandquist et al. 2003a;
2: van den Berg et al. 2001; 3: Sandquist \& Shetrone 2003c; 4:
Mathieu, Latham, \& Griffin 1990; 5: Sandquist \& Shetrone 2003b; 6:
Gilliland et al. 1991; 7: van den Berg et al. 2002; 8: Mathieu et
al. 2003; 9: Shetrone \& Sandquist 2000; 10: Belloni et al. 1998 
\end{table}

Elimination of known binaries will of course not eliminate all such
systems due to observational limitations (particularly for the radial
velocity surveys) and to selection effects (like those against
binaries with low mass ratios or long periods). Thus, it is necessary
to fall back on CMD position selection.  Because of the high relative
precision of the photometry for most stars in our study, we are able to
eliminate from consideration stars that are likely to be unresolved
binaries. However, as these cuts may not be restrictive enough in
removing unresolved binaries with the lowest mass ratios, we have also
used the data of \citet{fan} for an additional cut. Since faint
secondary stars can have significant effects on the
color of a system in spite of faintness, we have examined the position
of our selected stars in the CMD using several color
indices. A significant shift in position relative to other stars near
the SSS is an indication of an unresolved secondary star. This method
works best if measurement scatter is small, which is happily the
case for a subset of the filter bands used in the \citet{fan} study
(specifically, the filters centered at 3890, 6075, and 9190 \AA).

\begin{figure}
\includegraphics[width=84mm]{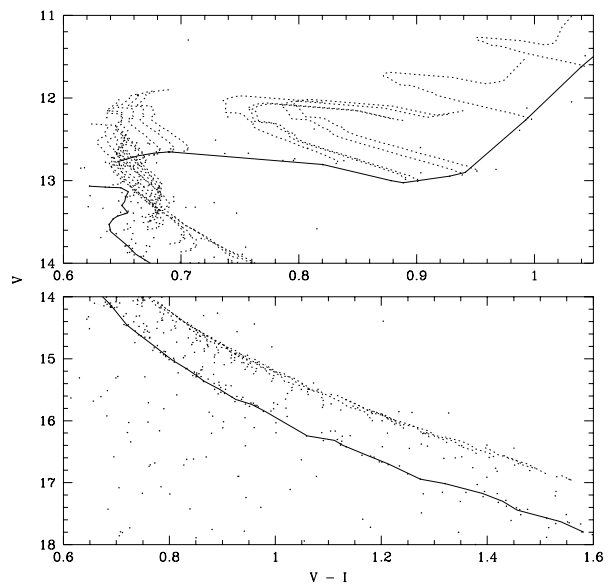}
\caption{CMDs showing our derived fiducial line ({\it solid line}),
two-star blend sequences for many different single star primaries ({\it dotted
lines}), and our cluster photometry ({\it points}).
\label{blends}}
\end{figure}

In different portions of the CMD, slightly different criteria were
used to select the stars that were most likely to be part of the SSS.
These criteria were based on the direction a single star is likely to
be displaced in the CMD if another single star was placed in an
unresolved binary with it.  Fig. \ref{blends} shows examples of this
for various places on the $VI$ fiducial line. To create this plot, we
combine the $V$ and $I$ fluxes of a given fiducial point with that of
each fainter fiducial point in order to determine where unresolved
binaries would fall. The indication is that for all points up to the
bottom of the turnoff gap ($12.9 < V < 13.3$) a ``bluest member''
selection criterion is appropriate since there is no possibility of an
unresolved binary falling exactly on the SSS. For all but the bluest
portion of the subgiant branch, a ``faintest member'' criterion is
appropriate. For the red giant branch, a ``reddest member'' criterion
works except on the lower RGB where it is possible to have a very
slight red increment added if a secondary star is among the brightest
main sequence stars redder than the giant primary. (We have found at
least one example of this in our dataset; see \S \ref{lrgb}) So, it is
only the bluest portion of the subgiant branch where there is likely
to be significant contamination from unresolved blends with main
sequence stars.

\begin{figure}
\includegraphics[width=84mm]{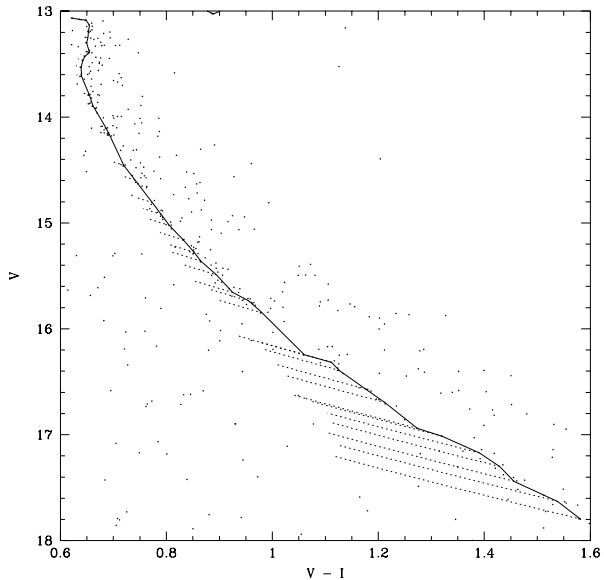}
\caption{CMDs showing our derived fiducial line ({\it solid line}),
two star blend sequences for unseen white dwarf companions ({\it dotted lines})
using the sequences of Bergeron, Wesemael, \& Beauchamp (1995), and
our cluster photometry ({\it points}).
\label{blendswd}}
\end{figure}

It is also possible that white dwarfs might be present in binary
stars, which would result in a displacement of the system's properties
to the blue of the main sequence as shown in Fig. \ref{blendswd}. The
white dwarf photometry values were taken from the $\log g = 8$ pure
hydrogen atmosphere models of Bergeron, Wesemael, \& Beauchamp (1995)
for $T_{\mbox{eff}} \le 10^5$ K. The displacement only begins to become
important for main sequence stars fainter than the turnoff. Blends
with white dwarfs should have negligible effects on the determined
position of the lower main sequence as they would most likely be
rejected as apparent field stars on the basis of their CMD position.
If they are common enough in binaries with stars on the upper main
sequence, they would produce systems in the CMD that appear to be
slightly ($\sim 0.01$ mag in color) to the blue of the main
sequence. This is one good reason for retaining some skepticism about
using a ``bluest member'' selection criterion here.

Strong interactions between stars can produce systems that do not
follow blend sequences. Blue stragglers are a notable example of a
group of stars that can fall between the SSS and the zero-age main
sequence because the hydrogen content in the stellar core is larger
than expected for an isolated single star of near-turnoff mass. The
current leading explanations for blue stragglers are earlier episodes
of binary mass transfer or stellar collisions. Blue stragglers that
fall near the turnoff in the CMD can be difficult to remove. In some
cases a second star will be present that still contributes enough to
the system flux (whether a leftover donor star or a star that was
captured into a bound orbit during a multiple star interaction) that
the star will be clearly separated from the single-star sequence. In
other cases, radial velocity studies \citep{mathieurv} can eliminate
some systems even if the secondary is undetectable on its own.

Three examples of interacting W UMa variable stars fall near the M67
turnoff, but it is unlikely that stars of this type will be a
significant contaminant because they are usually easily identified
from photometric studies like the ones carried out by \citet{SvMV} and
\citet{sandb} unless the system has a low inclination. Such systems
generally will not have system photometry that moves them bluer than
the single-star sequence \citep{rucdue} unless one of the stars was
already a blue straggler when strong interactions began. However, we
do have to be aware that there is a small probability that a blue
straggler or interacting star could fall near the SSS.

In a small number of cases we have also eliminated stars that appear
to be low amplitude variable stars. With a large number of
observations, it should in principle be possible to determine an
accurate median magnitude for stars that vary on fairly short
time-scales ($P \la 10$ d). In some cases variability will be related
to binary interactions or to rotational modulation of star spots. In
either case, the variability can introduce enough uncertainty that it
could affect the position of the star in the CMD. We have used our
time-series photometry and the study of \citet{SvMV} to identify and
remove these stars. We confirm low amplitude variability of the stars
S974 and S1093 detected by Stassun et al., and in addition find that
S1042 also appears to be variable (although Sanders 1977 gives it a
0\% membership probability). In addition, although S1112 made it
through our other cuts, it was shown in \citet{vSVM} to be a
low-amplitude variable, and so was eliminated.

\subsection{Other Means of Broadening the SSS\label{broad}}

In using the above criteria to select single stars, we may introduce 
some bias into our sample because other factors can lead to
a broadening of the region of the CMD containing single stars. Factors
include differences in age, chemical composition, and angular momentum.
We will briefly discuss these below.

For a cluster as old as M67, differences in age due to variations in
the time of formation result in relatively small differences in CMD
position.  The effects are maximized for the most evolved stars in the
cluster. Our selection criteria would tend to bias us toward selecting
the oldest stars in the cluster on the upper main sequence, SGB, and
RGB.  Age spreads in young open clusters are typically found to be
less than 10 Myr (e.g. Belikov et al. 2000; Soderblom et al. 1999;
Baume, V\'{a}zquez, \& Feinstein 1999). From isochrones (Yi et
al. 2001), we find that such an age difference would cause a color
shift of only about 0.001 in $V-I$ at the turnoff, and about 0.004 mag
in $V$ on the subgiant branch. On this basis, we neglect further
discussion of age spreads.

Composition differences can also create differences in properties
among stars of the same age and mass. There are no large spectroscopic
studies of M67 to give us an accurate measure of how much of an
intrinsic scatter in [Fe/H] there might be in the cluster. Indications
from 4 stars observed by Shetrone \& Sandquist (2000) are that the
scatter among main sequence stars is less than 0.10 dex. Observations
of 16 dwarfs in the Pleiades by Wilden et al. (2002) indicated
deviations from the mean of less than 0.03 dex. At the turnoff, a 0.1
dex difference in [Fe/H] would result in a small change in the shape of
the sequence, a change in turnoff color by a little over 0.01, and a
change in $V$ by almost 0.1 mag. The slope of the SGB is also
affected, resulting in the largest differences at the red end of
around 0.2 mag in $V$.  Our selection criteria would tend to be biased
toward the stars with the lowest metallicity values on the main
sequence, but the highest metallicity stars on the SGB and
RGB. Although this bias could potentially affect the shape of single
star sequence, we will not discuss this further because there is as
yet no evidence of significant metallicity scatter. We will discuss
the possibility of evolution-driven composition effects in \S
\ref{lrgb}.

Star-to-star angular momentum differences are believed to affect main
sequence Li abundances (e.g. Jones, Fischer, \& Soderblom 1999),
although the most abundant chemical species are not significantly
affected. Changes in CMD position on the main sequence are thus not
likely to result from the composition changes. Observations of
$v_{rot} \sin i$ for M67 stars show values of less than 8 km s$^{-1}$
(Melo et al. 2001). These low rotation rates are not likely to affect
the stellar structure enough to move the star significantly in the
CMD. As a result, we will drop further discussion of angular momentum
effects.

\subsection{Selected Stars and the SSS}

We determined main sequence fiducial points using our calibrated
ensemble photometry.  In places (particularly on the subgiant branch
and above), individual stars were used to define the SSS if we deemed
that there was sufficient evidence to support this. In places where
there were more than one star clumped together in the CMD, calculated
fiducial points were weighted averages of photometric values for the
group.  Our fiducial points are presented in Table \ref{fiduc},
including the number of stars $N$ used to compute each point.

\begin{table}
\caption{VI Fiducial Points for M67}
\label{fiduc}
\begin{tabular}{@{}ccc}
\hline
$V$ & $V-I$ & $N$ \\
\hline
11.4291 & 1.0557 & 1 \\
11.6221 & 1.0400 & 1 \\
12.2669 & 0.9926 & 1 \\
12.9001 & 0.9414 & 1 \\
12.9475 & 0.9281 & 1 \\
13.0265 & 0.8886 & 1 \\
13.0015 & 0.8782 & 1 \\
12.8059 & 0.8203 & 1 \\
12.6539 & 0.6899 & 2 \\
12.6677 & 0.6768 & 3 \\
12.7012 & 0.6634 & 2 \\
12.7677 & 0.6470 & 3 \\
12.8189 & 0.6479 & 4 \\
13.0667 & 0.6222 & 1 \\
13.0791 & 0.6370 & 2 \\
13.0861 & 0.6487 & 1 \\
13.1333 & 0.6552 & 3 \\
13.1907 & 0.6530 & 3 \\
13.2516 & 0.6520 & 2 \\
13.3012 & 0.6497 & 1 \\
13.3908 & 0.6549 & 5 \\
13.4271 & 0.6462 & 1 \\
13.4651 & 0.6423 & 1 \\
13.5305 & 0.6388 & 1 \\
13.6178 & 0.6402 & 3 \\
13.7941 & 0.6549 & 4 \\
13.8925 & 0.6616 & 7 \\
14.1064 & 0.6862 & 4 \\
14.1657 & 0.6925 & 6 \\
14.4573 & 0.7199 & 3 \\
14.5950 & 0.7407 & 3 \\
14.8078 & 0.7736 & 3 \\
14.9184 & 0.7902 & 5 \\
15.0238 & 0.8057 & 4 \\
15.1593 & 0.8319 & 2 \\
15.2770 & 0.8526 & 6 \\
15.3586 & 0.8648 & 3 \\
15.4896 & 0.8943 & 9 \\
15.6565 & 0.9250 & 3 \\
15.7368 & 0.9558 & 8 \\
15.8492 & 0.9803 & 2 \\
16.2449 & 1.0603 & 2 \\
16.3165 & 1.1117 & 3 \\
16.3931 & 1.1265 & 3 \\
16.5692 & 1.1762 & 3 \\
16.6970 & 1.2128 & 4 \\
16.9410 & 1.2737 & 1 \\
17.0145 & 1.3203 & 1 \\
17.1726 & 1.3907 & 4 \\
17.3017 & 1.4294 & 2 \\
17.4407 & 1.4562 & 5 \\
17.6307 & 1.5382 & 5 \\
17.7996 & 1.5821 & 3 \\
\hline
\end{tabular}
\end{table}

We truncate our tabulated sequence at $V \approx 17.8$ because
this corresponds to a color at the red end of the range covered by our
calibrating secondary standards. However, we are able to identify
additional stars that form an extension of the main sequence.  Our
selected stars are presented in Table \ref{sss}. Columns 1 -- 3
contain identification numbers from \citet{sanders}, \citet{mmj}, and
\citet{fan}, columns 4 -- 9 contain the photometry and numbers of
observations in each filter band, and columns 10 and 11 contain proper
motion membership probabilities from \citet{sanders} and
\citet{girard}. (Note that we tabulate magnitudes and colors to
fractions of millimag because in many cases the {\it relative}
accuracies justify this. The possibility of calibration errors means
that these could deviate from the true photometry values on an {\it
absolute} scale.) 
A subset of stars had questionable values resulting
from small numbers of measurements, contamination due to light from
nearby stars, or instrumental problems. If the star falls near the
apparent position of the SSS we include it in the table for possible
future use, and note it with an ``$a$''. We are confident in our choices
for the SSS based on the degree to which stars in our data and that of
\citet{fan} fall in the same regions of the CMD, as can be seen in
Figs.  \ref{rgbsgb}, \ref{cmdto}, and \ref{cmdms} below.

\begin{table*}
\centering
\begin{minipage}{15cm}
\caption{High-Probability Single Stars in M67}
\label{sss}
\begin{tabular}{@{}ccccrcrcrrrl@{}}
\hline
Sanders & MMJ & Fan & $V$ & $N_{V}$ & $V-I$ & $N_{I}$ & $B-V$ & 
$N_{B}$ & $P_{S}$ & $P_{G}$ & Notes \\
\hline
\multicolumn{12}{c}{\underline{Clump Stars}} \\
1010 & 6485 & 3536 & $10.4675\pm0.0005$ & 1711 & $1.0737\pm0.0020$ & 190 & $1.0946\pm0.0032$ & 62 & 0.96 & 0.99 & \\
1074 & 6492 & 3204 & $10.5221\pm0.0005$ & 1991 & $1.0721\pm0.0019$ & 144 & $1.0924\pm0.0019$ & 62 & 0.91 & 0.98 & \\
1084 & 6494 & 3642 & $10.4929\pm0.0004$ & 2158 & $1.0547\pm0.0020$ & 134 & $1.0855\pm0.0022$ & 62 & 0.92 & 0.98 & \\
1279 & 6503 & 3726 & $10.5379\pm0.0003$ & 1814 & $1.0859\pm0.0012$ & 191 & $1.1144\pm0.0020$ & 62 & 0.92 & 0.99 & \\
1316 & 6506 & 4159 &                    &      &                   &     &                   &    & 0.99 & 0.95 & \\
1479 & 6512 & 4611 &                    &      &                   &     &                   &    & 0.98 & 0.95 & \\
1592 & 6516 & 5050 &                    &      &                   &     &                   &    & 0.97 & 0.92 & \\ 
\multicolumn{12}{c}{\underline{Lower Giant Branch}} \\
989  &      & 3492 & $11.4291\pm0.0003$ & 2148 & $1.0557\pm0.0006$ & 564 & $1.0702\pm0.0199$ & 62 & 0.95 & 0.99 & \\
1277 & 6502 & 4117 & $11.6221\pm0.0005$ & 1240 & $1.0400\pm0.0006$ & 392 & $1.0367\pm0.0025$ & 62 & 0.95 & 0.98 & \\
1305 & 5997 & 3949 & $12.2669\pm0.0003$ & 2446 & $0.9926\pm0.0005$ & 668 & $0.9812\pm0.0018$ & 62 & 0.95 & 0.99 & \\
1231 & 5855 & 3736 & $12.9001\pm0.0004$ & 2150 & $0.9414\pm0.0005$ & 815 & $0.9179\pm0.0035$ & 62 & 0.95 & 0.98 & \\
606  & 5059 & 2210 &                    &      &                   &     &                   &    & 0.96 & 0.98 & \\
1103 & 5663 & 3413 &                    &      &                   &     &                   &    & 0.95 & 0.99 & \\
1402 & 6508 & 4878 &                    &      &                   &     &                   &    & 0.83 & 0.77 & \\
1585 &      & 5191 &                    &      &                   &     &                   &    & 0.89 & 0.91 & \\ 
\multicolumn{12}{c}{\underline{Subgiant Branch}} \\
1245 & 6114 & 4178 & $12.9475\pm0.0008$ &  922 & $0.9281\pm0.0012$ & 362 & $0.8925\pm0.0031$ & 56 & 0.95 & 0.98 & $a$ \\
1060 & 5651 & 3406 & $13.0265\pm0.0006$ & 2449 & $0.8886\pm0.0007$ & 819 & $0.8545\pm0.0023$ & 62 & 0.96 & 0.98 & $a$ \\
1056 & 5580 & 3296 & $13.0015\pm0.0005$ & 2394 & $0.8782\pm0.0006$ & 820 & $0.8379\pm0.0027$ & 62 & 0.95 & 0.96 & $a$ \\
1323 & 5996 & 3935 & $12.8059\pm0.0012$ &  491 & $0.8203\pm0.0086$ &   8 &                 &  0 & 0.95 & 0.99 & $a$ \\
756  & 5388 & 2932 & $12.6546\pm0.0007$ & 2015 & $0.6903\pm0.0009$ & 810 & $0.6128\pm0.0319$ & 62 & 0.96 & 0.99 & $a$ \\
775  & 5371 & 2887 & $12.6526\pm0.0010$ & 1877 & $0.6892\pm0.0011$ & 818 & $0.6130\pm0.0025$ & 62 & 0.84 & 0.99 & $a$ \\
1034 & 5644 & 3395 & $12.6530\pm0.0003$ & 2011 & $0.6790\pm0.0004$ & 819 & $0.6056\pm0.0011$ & 62 & 0.95 & 0.99 & $a$ \\
1270 & 6166 & 4314 & $12.6843\pm0.0013$ &  589 & $0.6771\pm0.0034$ & 181 & $0.5578\pm0.0069$ & 27 & 0.94 & 0.99 & \\
2212 & 5884 & 3781 & $12.6922\pm0.0004$ & 2011 & $0.6728\pm0.0006$ & 819 & $0.5869\pm0.0016$ & 62 & 0.95 & 0.99 & \\
598  & 5041 & 2172 &                    &      &                   &     &                   &    & 0.96 & 0.99 & \\
859  &      & 2481 &                    &      &                   &     &                   &    & 0.89 & 0.97 & \\
1268 & 6177 & 4335 &                    &      &                   &     &                   &    & 0.94 & 0.99 & \\
1274 & 5925 & 3836 &                    &      &                   &     &                   &    & 0.93 & 0.99 & \\
\multicolumn{12}{c}{\underline{The Turnoff}} \\
1456 & 6224 & 4434 & $12.7160\pm0.0008$ &  335 & $0.6619\pm0.0023$ & 106 & $0.5181\pm0.0074$ & 25 & 0.94 & 0.99 & \\
     & 5951 & 3874 & $12.6825\pm0.0009$ & 1593 & $0.6637\pm0.0010$ & 798 & $0.5806\pm0.0039$ & 62 &     &     & $a$ \\
995  & 5675 & 3454 & $12.7702\pm0.0002$ & 2015 & $0.6463\pm0.0004$ & 815 & $0.5501\pm0.0027$ & 62 & 0.82 & 0.99 & $a$ \\
1083 & 5825 & 3686 & $12.7547\pm0.0004$ & 2448 & $0.6495\pm0.0005$ & 672 & $0.5678\pm0.0016$ & 62 & 0.95 & 0.99 & \\
1310 & 6077 & 4091 & $12.7792\pm0.0008$ & 1510 & $0.6358\pm0.0016$ & 478 & $0.5601\pm0.0026$ & 61 & 0.93 & 0.99 & \\
1003 & 5562 & 3267 & $12.8281\pm0.0005$ & 2008 & $0.6407\pm0.0007$ & 814 & $0.5550\pm0.0032$ & 62 & 0.91 & 0.99 & $a$ \\
1049 & 5573 & 3283 & $12.8098\pm0.0005$ & 2045 & $0.6422\pm0.0006$ & 816 & $0.5552\pm0.0029$ & 62 & 0.95 & 0.98 & $a$ \\
1071 & 5842 & 3708 & $12.8111\pm0.0003$ & 2436 & $0.6511\pm0.0006$ & 673 & $0.5639\pm0.0037$ & 62 & 0.94 & 0.98 & \\
1076 & 5586 & 3302 & $12.8266\pm0.0003$ & 2416 & $0.6525\pm0.0005$ & 669 & $0.5571\pm0.0016$ & 62 & 0.94 & 0.98 & \\
926  &      & 3632 &                    &      &                   &     &                   &    & 0.94 & 0.98 & \\
1589 &      & 5052 &                    &      &                   &     &                   &    & 0.94 & 0.98 & \\
1639 &      & 5318 &                    &      &                   &     &                   &    & 0.90 & 0.95 & \\
\multicolumn{12}{c}{\underline{Lower Turnoff}} \\
731  & 5335 & 2841 &                    &      &                   &     &                   &    & 0.96 & 0.98 & \\
1230 & 6103 & 4153 & $13.0667\pm0.0016$ & 1219 & $0.6222\pm0.0018$ & 470 & $0.5364\pm0.0028$ & 52 & 0.94 & 0.99 & $a$ \\
998  & 5610 & 3359 & $13.0803\pm0.0004$ & 2013 & $0.6389\pm0.0006$ & 816 & $0.5522\pm0.0027$ & 62 & 0.95 & 0.99 & $a$ \\
1302 & 5926 & 3832 & $13.0779\pm0.0004$ & 2403 & $0.6356\pm0.0006$ & 669 & $0.5605\pm0.0037$ & 62 & 0.95 & 0.97 & $a$ \\
745  & 5169 & 2461 & $13.0861\pm0.0007$ &  849 & $0.6487\pm0.0041$ &  29 &                 &  0 & 0.96 & 0.99 & $a$ \\
827  & 5312 & 2765 &                    &      &                   &     &                   &    & 0.95 & 0.95 & \\
1219 & 5863 & 3766 & $13.1130\pm0.0005$ & 1703 & $0.6590\pm0.0008$ & 747 & $0.5725\pm0.0020$ & 61 & 0.93 & 0.99 & \\
2205 & 5679 & 3458 & $13.1453\pm0.0004$ & 2152 & $0.6498\pm0.0008$ & 809 & $0.5642\pm0.0056$ & 62 & 0.51 & 0.98 & $a$ \\
2220 & 5597 & 3335 & $13.1424\pm0.0012$ & 2041 & $0.6603\pm0.0016$ & 812 & $0.5882\pm0.0025$ & 62 & 0.04 & 0.99 & \\
994  & 5716 & 3510 & $13.1956\pm0.0003$ & 2012 & $0.6526\pm0.0006$ & 810 & $0.5678\pm0.0024$ & 62 & 0.95 & 0.98 & \\
1207 & 6108 & 4174 & $13.1763\pm0.0005$ &  744 & $0.6597\pm0.0031$ &  91 & $0.5349\pm0.0561$ & 13 & 0.95 & 0.98 & \\
1313 & 6019 & 3977 & $13.1985\pm0.0018$ & 1430 & $0.6542\pm0.0019$ & 371 & $0.5660\pm0.0065$ & 56 & 0.94 & 0.98 & \\
1181 &      & 4119 &                    &      &                   &     &                   &    & 0.96 & 0.99 & \\
1197 & 5959 & 3909 &                    &      &                   &     &                   &    & 0.95 & 0.99 & \\
1441 & 6336 & 4689 &                    &      &                   &     &                   &    & 0.95 & 0.98 & \\
\end{tabular}
\end{minipage}
\end{table*}

\begin{table*}
\begin{minipage}{15cm}
\contcaption{}
\begin{tabular}{@{}ccccrcrcrrrl@{}}
\hline
Sanders & MMJ & Fan & $V$ & $N_{V}$ & $V-I$ & $N_{I}$ & $B-V$ & 
$N_{B}$ & $P_{S}$ & $P_{G}$ & Notes \\
\hline
964  & 5622 & 3383 & $13.2544\pm0.0004$ & 1871 & $0.6574\pm0.0009$ & 613 & $0.5826\pm0.0038$ & 49 & 0.95 & 0.97 & \\
1030 & 5803 & 3633 & $13.2471\pm0.0005$ & 2007 & $0.6470\pm0.0009$ & 814 & $0.5674\pm0.0028$ & 62 & 0.90 & 0.98 & $a$ \\
1635 &      & 5352 &                    &      &                   &     &                   &    & 0.71 & 0.54 & \\
1062 & 5733 & 3527 & $13.3012\pm0.0004$ & 2445 & $0.6497\pm0.0006$ & 771 & $0.5646\pm0.0014$ & 62 & 0.96 & 0.97 & \\
1189 & 5900 & 3838 &                    &      &                   &     &                   &    & 0.95 & 0.97 & \\
1458 & 6332 & 4676 &                    &      &                   &     &                   &    & 0.95 & 0.97 & \\
648  &      & 2365 &                    &      &                   &     &                   &    & 0.93 & 0.73 & \\
711  &      & 2653 &                    &      &                   &     &                   &    & 0.90 & 0.96 & \\
723  & 5277 & 2726 &                    &      &                   &     &                   &    & 0.87 & 0.93 & \\
809  & 5314 & 2778 & $13.3839\pm0.0006$ &  997 & $0.6486\pm0.0033$ & 239 & $0.5545\pm0.0039$ & 25 & 0.94 & 0.96 & $a$ \\ 
967  & 5629 & 3390 & $13.3986\pm0.0004$ & 1784 & $0.6525\pm0.0006$ & 742 & $0.5673\pm0.0040$ & 49 & 0.94 & 0.99 & \\
1017 & 5478 & 3102 & $13.3933\pm0.0006$ & 1928 & $0.6595\pm0.0008$ & 812 & $0.5504\pm0.0036$ & 62 & 0.95 & 0.97 & \\
1240 & 6134 & 4230 & $13.3698\pm0.0011$ &  778 & $0.6550\pm0.0020$ & 322 & $0.5537\pm0.0140$ & 44 & 0.93 & 0.98 & \\
2221 & 5559 & 3254 & $13.3878\pm0.0004$ & 2157 & $0.6544\pm0.0006$ & 813 & $0.5742\pm0.0019$ & 62 & 0.90 & 0.98 & \\
839  & 5183 & 2463 &                    &      &                   &     &                   &    & 0.85 & 0.84 & \\
939  & 5434 & 3038 &                    &      &                   &     &                   &    & 0.95 & 0.99 & \\
990  & 5464 & 3083 & $13.4271\pm0.0005$ & 2019 & $0.6462\pm0.0009$ & 811 & $0.5507\pm0.0016$ & 53 & 0.93 & 0.97 & $a$ \\
740  & 5354 & 2881 & $13.4651\pm0.0003$ & 1757 & $0.6423\pm0.0009$ & 365 & $0.5476\pm0.0070$ & 25 & 0.87 & 0.98 & $a$ \\
784  & 5283 & 2698 & $13.5305\pm0.0006$ & 1906 & $0.6388\pm0.0011$ & 519 & $0.5491\pm0.0047$ & 35 & 0.91 & 0.99 & $a$ \\
1321 & 6127 & 4192 & $13.5674\pm0.0021$ &  203 & $0.6569\pm0.0065$ &  28 & $0.5487\pm0.0321$ & 13 & 0.94 & 0.98 & $b$ \\
763  & 5196 & 2494 & $13.5966\pm0.0005$ &  799 & $0.6394\pm0.0024$ & 143 &                 &  0 & 0.95 & 0.98 & \\
803  & 5301 & 2733 & $13.6420\pm0.0005$ & 1954 & $0.6449\pm0.0009$ & 541 & $0.5635\pm0.0020$ & 37 & 0.95 & 0.99 & \\
1061 & 5546 & 3227 & $13.6147\pm0.0005$ & 2428 & $0.6370\pm0.0008$ & 812 & $0.5419\pm0.0044$ & 62 & 0.96 & 0.97 & $a$ \\
844  &      & 2526 &                    &      &                   &     &                   &    & 0.92 & 0.97 & \\
\multicolumn{12}{c}{\underline{Upper Main Sequence}} \\
1115 &      & 3097 &                    &      &                   &     &                   &    & 0.91 & 0.95 & \\
1281 & 5972 & 3903 & $13.6886\pm0.0007$ & 2211 & $0.6318\pm0.0010$ & 765 & $0.5272\pm0.0049$ & 60 & 0.90 & 0.99 & BSS?\\
595  &      & 2039 &                    &      &                   &     &                   &    & 0.94 & 0.90 & \\
1055 & 5471 & 3080 & $13.7936\pm0.0007$ & 2298 & $0.6523\pm0.0010$ & 811 & $0.5679\pm0.0029$ & 62 & 0.96 & 0.98 & $a$ \\
1244 & 6058 & 4071 & $13.8197\pm0.0008$ & 1433 & $0.6586\pm0.0010$ & 643 & $0.5807\pm0.0049$ & 61 & 0.92 & 0.97 & \\
1265 & 6065 & 4077 & $13.7863\pm0.0006$ & 1277 & $0.6532\pm0.0008$ & 636 & $0.5697\pm0.0023$ & 62 & 0.94 & 0.98 & \\
1300 & 6060 & 4057 & $13.7857\pm0.0007$ & 1898 & $0.6559\pm0.0009$ & 648 & $0.5791\pm0.0033$ & 62 & 0.94 & 0.95 & \\
755  & 5189 & 2491 & $13.8996\pm0.0008$ &  791 & $0.6601\pm0.0020$ & 138 &                 &  0 & 0.96 & 0.98 & $a$ \\
796  & 5412 & 2978 & $13.8526\pm0.0005$ & 2304 & $0.6583\pm0.0007$ & 661 & $0.5762\pm0.0028$ & 62 & 0.92 & 0.98 & \\
987  & 5608 & 3360 & $13.9182\pm0.0004$ & 2139 & $0.6659\pm0.0008$ & 805 & $0.5917\pm0.0082$ & 61 & 0.95 & 0.98 & \\
1021 & 5522 & 3186 & $13.9133\pm0.0008$ & 1951 & $0.6680\pm0.0010$ & 805 & $0.5770\pm0.0023$ & 62 & 0.89 & 0.98 & \\
1035 & 5657 & 3425 & $13.8677\pm0.0005$ & 1994 & $0.6563\pm0.0011$ & 810 & $0.5757\pm0.0033$ & 62 & 0.90 & 0.98 & \\
1051 & 5685 & 3455 & $13.9627\pm0.0005$ & 2031 & $0.6693\pm0.0009$ & 805 & $0.5811\pm0.0020$ & 62 & 0.94 & 0.96 & $a$ \\
1075 & 5692 & 3456 & $13.8722\pm0.0003$ & 2293 & $0.6544\pm0.0008$ & 661 & $0.5682\pm0.0037$ & 62 & 0.93 & 0.97 & $a$ \\
647  &      & 2301 &                    &      &                   &     &                   &    & 0.95 & 0.96 & \\
650  &      & 1902 &                    &      &                   &     &                   &    & 0.92 & 0.75 & \\
1201 & 5989 & 3954 &                    &      &                   &     &                   &    & 0.94 & 0.98 & \\
848  &      & 2769 &                    &      &                   &     &                   &    & 0.95 & 0.93 & \\
1318 &      & 3960 &                    &      &                   &     &                   &    & 0.86 & 0.90 & \\
805  & 5229 & 2559 & $14.0073\pm0.0006$ & 1107 & $0.6812\pm0.0018$ & 187 &                 &  0 & 0.95 & 0.93 & $b$ \\
1446 & 6395 & 4842 &                    &      &                   &     &                   &    & 0.94 & 0.74 & \\
1406 & 6237 & 4479 &                    &      &                   &     &                   &    & 0.85 & 0.96 & \\
1330 & 6104 & 4132 &                    &      &                   &     &                   &    & 0.79 & 0.82 & \\
603  & 5111 & 2305 &                    &      &                   &     &                   &    & 0.93 & 0.97 & \\
744  & 5181 & 2484 & $14.1219\pm0.0009$ &  860 & $0.6824\pm0.0026$ & 110 &                  & 0 & 0.95 & 0.79 & $a$ \\
788  & 5305 & 2746 & $14.1117\pm0.0008$ & 1947 & $0.6867\pm0.0012$ & 539 & $0.6077\pm0.0030$ & 37 & 0.95 & 0.97 & \\
1213 & 6031 & 4020 & $14.0869\pm0.0015$ & 1857 & $0.6791\pm0.0017$ & 466 & $0.5995\pm0.0057$ & 40 & 0.96 & 0.98 & $a$ \\
1283 & 5873 & 3758 & $14.1018\pm0.0005$ & 2396 & $0.6888\pm0.0010$ & 801 & $0.6207\pm0.0030$ & 60 & 0.96 & 0.96 & \\
722  & 5307 & 2781 &                    &      &                   &     &                   &    & 0.96 & 0.90 & \\
1107 & 5722 & 3491 &                    &      &                   &     &                   &    & 0.73 & 0.96 & \\
982  & 5776 & 3608 & $14.1056\pm0.0004$ & 2132 & $0.6586\pm0.0007$ & 784 & $0.5943\pm0.0050$ & 60 & 0.93 & 0.94 & $b$ \\
1033 & 5567 & 3274 & $14.1553\pm0.0005$ & 1993 & $0.6888\pm0.0008$ & 803 & $0.6039\pm0.0051$ & 62 & 0.91 & 0.97 & $a$ \\
1078 & 5826 & 3687 & $14.1781\pm0.0005$ & 2391 & $0.6966\pm0.0009$ & 660 & $0.6281\pm0.0046$ & 61 & 0.85 & 0.97 & \\
1087 & 5753 & 3551 & $14.1697\pm0.0006$ & 2272 & $0.6888\pm0.0008$ & 658 & $0.6217\pm0.0034$ & 58 & 0.79 & 0.84 & $a$ \\
1089 & 5532 & 3191 & $14.1695\pm0.0006$ & 1799 & $0.7010\pm0.0008$ & 525 & $0.6217\pm0.0023$ & 59 & 0.91 & 0.91 & \\
1248 & 5963 & 3896 & $14.1694\pm0.0004$ & 1988 & $0.6903\pm0.0006$ & 758 & $0.6216\pm0.0045$ & 60 & 0.93 & 0.96 & \\
\end{tabular}
\end{minipage}
\end{table*}

\begin{table*}
\begin{minipage}{15cm}
\contcaption{}
\begin{tabular}{@{}ccccrcrcrrrl@{}}
\hline
Sanders & MMJ & Fan & $V$ & $N_{V}$ & $V-I$ & $N_{I}$ & $B-V$ & 
$N_{B}$ & $P_{S}$ & $P_{G}$ & Notes \\
\hline
1260 & 5937 & 3855 & $14.1471\pm0.0006$ & 1980 & $0.6909\pm0.0009$ & 799 & $0.6117\pm0.0056$ & 62 & 0.91 & 0.97 & \\
728  & 5294 & 2752 &                    &      &                   &     &                   &    & 0.94 & 0.76 & \\
937  & 5426 & 3031 &                    &      &                   &     &                   &    & 0.96 & 0.95 & \\
1505 &      & 4656 &                    &      &                   &     &                   &    & 0.94 & 0.79 & \\
630  &      & 2024 &                    &      &                   &     &                   &    & 0.96 & 0.91 & \\
1426 & 6405 & 4887 &                    &      &                   &     &                   &    & 0.95 & 0.91 & \\
1102 & 5479 & 3092 &                    &      &                   &     &                   &    & 0.95 & 0.86 & \\
829  & 5338 & 2806 &                    &      &                   &     &                   &    & 0.94 & 0.91 & \\
1423 & 6241 & 4478 &                    &      &                   &     &                   &    & 0.89 & 0.88 & \\
1449 & 6265 & 4515 &                    &      &                   &     &                   &    & 0.85 & 0.96 & \\
958  &      & 3249 & $14.4493\pm0.0009$ &  874 & $0.7143\pm0.0021$ & 257 &                 &  0 & 0.92 & 0.96 & \\
966  & 5704 & 3503 & $14.4644\pm0.0006$ & 1651 & $0.7187\pm0.0010$ & 693 & $0.6607\pm0.0033$ & 46 & 0.87 & 0.77 & $a$ \\
1255 & 5995 & 3953 & $14.4525\pm0.0007$ & 1972 & $0.7224\pm0.0010$ & 719 & $0.6631\pm0.0056$ & 61 & 0.86 & 0.94 & \\
1096 & 5541 & 3210 & $14.4743\pm0.0115$ &  235 & $0.7573\pm0.0118$ &  28 &                 &  0 & 0.90 & 0.94 & $b$ \\
1258 &      & 4309 & $14.4788\pm0.0012$ &  584 & $0.7428\pm0.0020$ & 213 & $0.6323\pm0.0307$ & 24 & 0.96 & 0.92 & $b$ \\
621  & 5039 & 2146 &                    &      &                   &     &                   &    & 0.95 & 0.87 & \\
942  & 5444 & 3059 &                    &      &                   &     &                   &    & 0.95 & 0.90 & \\
1421 &      & 4446 &                    &      &                   &     &                   &    & 0.81 & 0.80 & \\
1616 & 6460 & 4983 &                    &      &                   &     &                   &    & 0.95 & 0.85 & \\
945  & 5744 & 3570 &                    &      &                   &     &                   &    & 0.94 & 0.93 & \\
724  &      & 3017 &                    &      &                   &     &                   &    & 0.96 & 0.93 & \\
1452 & 6347 & 4706 &                    &      &                   &     &                   &    & 0.95 & 0.86 & \\
753  & 5243 & 2625 & $14.6158\pm0.0005$ & 1328 & $0.7418\pm0.0012$ & 381 & $0.6790\pm0.0027$ & 34 & 0.46 & 0.93 & $a$ \\
770  & 5357 & 2867 & $14.6296\pm0.0010$ & 1829 & $0.7466\pm0.0013$ & 791 & $0.6748\pm0.0044$ & 60 & 0.93 & 0.88 & \\
1218 & 5966 & 3910 & $14.5525\pm0.0006$ & 1638 & $0.7357\pm0.0011$ & 682 & $0.6731\pm0.0043$ & 61 & 0.93 & 0.97 & \\
779  & 5306 & 2758 &                    &      &                   &     &                   &    & 0.93 & 0.85 & \\
951  & 5489 & 3141 &                    &      &                   &     &                   &    & 0.94 & 0.87 & \\
1184 &      & 3937 &                    &      &                   &     &                   &    & 0.94 & 0.87 & \\
2211 & 6387 & 4820 &                    &      &                   &     &                   &    & 0.58 & 0.76 & \\
1341 & 5908 & 3789 &                    &      &                   &     &                   &    & 0.83 & 0.94 & \\
1106 & 5469 & 3056 &                    &      &                   &     &                   &    & 0.87 & 0.91 & \\
629  & 5089 & 2251 &                    &      &                   &     &                   &    & 0.42 & 0.73 & \\
955  &      & 3309 &                    &      &                   &     &                   &    & 0.85 & 0.74 & \\
928  &      & 3104 &                    &      &                   &     &                   &    & 0.86 & 0.74 & \\
785  & 5346 & 2845 & $14.8244\pm0.0010$ & 2217 & $0.7715\pm0.0013$ & 736 & $0.7212\pm0.0073$ & 53 & 0.95 & 0.88 & $a$ \\
802  & 5360 & 2866 & $14.7875\pm0.0006$ & 2260 & $0.7774\pm0.0012$ & 651 & $0.7230\pm0.0076$ & 58 & 0.96 & 0.86 & \\
2213 & 5517 & 3181 & $14.8653\pm0.0012$ & 1903 & $0.7689\pm0.0018$ & 789 & $0.7105\pm0.0089$ & 59 & 0.95 & 0.57 & \\
640  &      & 1872 &                    &      &                   &     &                   &    & 0.51 & 0.63 & \\
754  & 5372 & 2901 & $14.9665\pm0.0011$ & 1923 & $0.7932\pm0.0014$ & 780 & $0.7431\pm0.0064$ & 59 & 0.83 & 0.47 & $a$ \\
1004 & 5768 & 3589 & $14.9470\pm0.0008$ & 1957 & $0.7929\pm0.0011$ & 790 & $0.7377\pm0.0055$ & 59 & 0.94 & 0.89 & \\
1067 & 5455 & 3043 & $14.9005\pm0.0007$ & 2287 & $0.7832\pm0.0011$ & 716 & $0.7253\pm0.0057$ & 58 & 0.60 & 0.62 & $a$ \\
1269 & 6080 & 4106 & $14.9361\pm0.0010$ & 1207 & $0.7892\pm0.0016$ & 567 & $0.7428\pm0.0050$ & 58 & 0.94 & 0.81 & $a$ \\
1289 & 5879 & 3768 & $14.8864\pm0.0007$ & 2376 & $0.7927\pm0.0011$ & 787 & $0.7513\pm0.0040$ & 60 & 0.90 & 0.94 & \\
820  & 5310 & 2767 & $14.9838\pm0.0020$ &  459 & $0.8266\pm0.0061$ &  26 &                 &  0 & 0.95 & 0.77 & $b$ \\
795  & 5391 & 2928 & $14.9935\pm0.0009$ & 2224 & $0.8081\pm0.0013$ & 647 & $0.7457\pm0.0059$ & 53 & 0.42 & 0.62 & \\
799  & 5297 & 2729 & $15.0205\pm0.0007$ & 1896 & $0.8091\pm0.0014$ & 518 & $0.7581\pm0.0048$ & 37 & 0.23 & 0.83 & \\ 
801  & 5424 & 3002 & $15.0563\pm0.0007$ & 2269 & $0.8095\pm0.0011$ & 641 & $0.7486\pm0.0101$ & 56 & 0.95 & 0.87 & \\
1068 & 5578 & 3287 & $15.0158\pm0.0006$ & 2354 & $0.7995\pm0.0010$ & 703 & $0.7469\pm0.0073$ & 53 & 0.24 & 0.83 & $a$ \\
772  & 5269 & 2663 & $15.1563\pm0.0007$ & 1353 & $0.8330\pm0.0015$ & 409 & $0.7677\pm0.0040$ & 35 & 0.91 & 0.49 & \\
1307 & 5949 & 3864 & $15.1615\pm0.0006$ & 2370 & $0.8310\pm0.0013$ & 642 & $0.7799\pm0.0096$ & 54 & 0.94 & 0.80 & \\
941  & 5771 & 3614 &                    &      &                   &     &                   &    & 0.94 & 0.69 & \\
1029 & 5709 & 3495 & $15.2151\pm0.0012$ & 1929 & $0.8472\pm0.0018$ & 778 & $0.7795\pm0.0096$ & 52 & 0.83 & 0.72 & \\
1079 & 5725 & 3505 & $15.2860\pm0.0008$ & 2308 & $0.8520\pm0.0020$ & 635 & $0.8073\pm0.0042$ & 51 & 0.94 & 0.63 & \\
1090 & 5526 & 3183 & $15.2722\pm0.0009$ & 1408 & $0.8444\pm0.0024$ & 397 & $0.7792\pm0.0050$ & 36 & 0.53 & 0.47 & $a$ \\
1217 & 6161 & 4311 & $15.2415\pm0.0021$ &  282 & $0.8558\pm0.0043$ & 147 &                 &  0 & 0.93 & 0.38 & \\
1227 & 6046 & 4046 & $15.2888\pm0.0008$ & 1776 & $0.8598\pm0.0013$ & 628 & $0.8208\pm0.0070$ & 51 & 0.00 & 0.32 & \\
1291 &      & 3926 & $15.2926\pm0.0008$ & 2326 & $0.8502\pm0.0014$ & 724 & $0.8150\pm0.0068$ & 57 & 0.94 & 0.78 & $a$ \\
1469 & 6321 & 4633 &                    &      &                   &     &                   &    & 0.94 & 0.23 & \\
626  &      & 1947 &                    &      &                   &     &                   &    & 0.67 & 0.42 & \\
1100 & 5560 & 3238 & $15.2300\pm0.0214$ &  213 & $0.8516\pm0.0219$ &  27 &                 &  0 & 0.81 & 0.61 & $b$ \\
1212 & 6139 & 4248 & $15.2550\pm0.0015$ &  643 & $0.8516\pm0.0137$ & 117 &                 &  0 & 0.63 & 0.83 & $b$ \\
2202 & 5551 & 3256 & $15.2289\pm0.0017$ &      & $0.8166\pm0.0032$ &     & $0.7787\pm0.0272$ &    & 0.74 & 0.47 & $b$ \\
\end{tabular}
\end{minipage}
\end{table*}

\begin{table*}
\begin{minipage}{15cm}
\contcaption{}
\begin{tabular}{@{}ccccrcrcrrrl@{}}
\hline
Sanders & MMJ & Fan & $V$ & $N_{V}$ & $V-I$ & $N_{I}$ & $B-V$ & 
$N_{B}$ & $P_{S}$ & $P_{G}$ & Notes \\
\hline
611  & 5090 & 2265 &                    &      &                   &     &                   &    & 0.92 & 0.76 & \\
778  & 5265 & 2657 & $15.3595\pm0.0007$ & 1340 & $0.8642\pm0.0015$ & 398 & $0.8108\pm0.0060$ & 34 & 0.00 & 0.28 & $a$ \\
804  & 5186 & 2477 & $15.3282\pm0.0018$ &  787 & $0.8731\pm0.0048$ & 110 &                 &  0 & 0.95 & 0.57 & \\
993  & 5441 & 3037 & $15.3678\pm0.0011$ & 1839 & $0.8647\pm0.0016$ & 774 & $0.8105\pm0.0059$ & 44 & 0.88 & 0.44 & $a$ \\
1028 & 5563 & 3268 & $15.3976\pm0.0008$ & 1920 & $0.8845\pm0.0013$ & 778 & $0.8326\pm0.0047$ & 55 & 0.00 & 0.66 & \\
1039 & 5459 & 3068 & $15.5666\pm0.0010$ & 1832 & $0.8921\pm0.0013$ & 770 & $0.8206\pm0.0083$ & 41 & 0.03 & 0.44 & \\
1081 & 5800 & 3623 & $15.5594\pm0.0007$ & 2312 & $0.9043\pm0.0016$ & 632 & $0.8550\pm0.0088$ & 47 & 0.00 & 0.66 & $a$ \\
1085 & 5539 & 3208 & $15.5411\pm0.0008$ & 2159 & $0.8972\pm0.0015$ & 629 & $0.8448\pm0.0096$ & 36 & 0.00 & 0.31 & $a$ \\
1215 & 5841 & 3720 & $15.4271\pm0.0009$ & 1633 & $0.9067\pm0.0014$ & 599 & $0.8496\pm0.0060$ & 40 & 0.67 & 0.08 & \\
1229 & 5904 & 3825 & $15.4645\pm0.0006$ & 2030 & $0.8971\pm0.0010$ & 775 & $0.8512\pm0.0051$ & 50 & 0.80 &      & \\
1304 & 5924 & 3829 & $15.4284\pm0.0009$ & 2338 & $0.8941\pm0.0015$ & 634 & $0.8461\pm0.0055$ & 46 & 0.71 & 0.21 & \\
1312 & 6015 & 3972 & $15.4940\pm0.0020$ & 1511 & $0.9061\pm0.0027$ & 353 & $0.8505\pm0.0092$ & 39 & 0.00 &      & \\
1443 &      & 4440 & $15.3794\pm0.0027$ &  281 & $0.8936\pm0.0047$ &  93 &                 &  0 & 0.94 & 0.16 & \\
2215 & 6005 & 3967 & $15.4949\pm0.0012$ & 1884 & $0.8855\pm0.0019$ & 681 & $0.8563\pm0.0117$ & 49 & 0.01 & 0.41 & $a$ \\
769  & 5319 & 2807 & $15.6086\pm0.0011$ & 1636 & $0.9275\pm0.0017$ & 613 & $0.8502\pm0.0048$ & 37 & 0.95 & 0.20 & \\
1038 & 5666 & 3438 & $15.6911\pm0.0009$ & 1885 & $0.9238\pm0.0014$ & 768 & $0.8620\pm0.0073$ & 43 & 0.00 & 0.17 & $a$ \\
1435 & 6214 & 4420 & $15.6206\pm0.0034$ &  182 & $0.9185\pm0.0046$ &  70 &                 &  0 & 0.00 &      & \\
     & 6102 & 4147 & $15.7126\pm0.0016$ & 1064 & $0.9540\pm0.0026$ & 453 & $0.9079\pm0.0109$ & 48 &      &     & \\
     & 5943 & 3861 & $15.7319\pm0.0014$ & 1706 & $0.9435\pm0.0021$ & 758 & $0.8641\pm0.0087$ & 38 &      &     & $a$ \\
764  & 5317 & 2804 & $15.7170\pm0.0010$ & 1600 & $0.9584\pm0.0018$ & 607 & $0.8966\pm0.0059$ & 35 & 0.00 &     & \\
811  & 5386 & 2914 & $15.7770\pm0.0017$ &  698 & $0.9676\pm0.0034$ & 273 & $0.9106\pm0.0185$ & 23 & 0.08 &     & \\
1091 & 5470 & 3069 & $15.7694\pm0.0036$ &  470 & $0.9512\pm0.0051$ & 232 & $0.8783\pm0.0133$ & 11 & 0.00 &     & $a$ \\
1098 & 5513 & 3161 & $15.7514\pm0.0068$ &  208 & $0.9614\pm0.0092$ &  26 &                 &  0 & 0.00 & 0.27 & \\
1311 & 5956 & 3867 & $15.7522\pm0.0011$ & 1540 & $0.9652\pm0.0025$ & 351 & $0.9076\pm0.0051$ & 34 & 0.00 &     & \\
1472 & 6248 & 4475 & $15.7653\pm0.0036$ &  166 & $0.9627\pm0.0102$ &  32 &                 &  0 & 0.00 &     & \\
960  & 5623 & 3389 & $15.8502\pm0.0015$ & 1206 & $0.9775\pm0.0024$ & 284 &                 &  0 & 0.00 &     & $a$ \\
1296 & 6195 & 4357 & $15.8443\pm0.0033$ &  469 & $0.9918\pm0.0050$ & 126 &                 &  0 & 0.00 &     & \\
1315 & 6096 & 4128 & $15.9311\pm0.0034$ &  424 & $1.0224\pm0.0077$ &  30 &                 &  0 & 0.00 &     & $b$ \\
1433 & 6229 & 4452 & $16.1932\pm0.0060$ &  138 & $1.0876\pm0.0096$ &  57 &                 &  0 & 0.00 &     & \\
     & 5754 & 3561 & $16.2477\pm0.0014$ & 1777 & $1.0587\pm0.0024$ & 745 & $0.9997\pm0.0131$ & 33 &      &     & \\
     & 5380 & 2911 & $16.3028\pm0.0016$ & 1618 & $1.1171\pm0.0022$ & 743 & $1.0250\pm0.0175$ & 32 &      &     & \\ 
     & 5501 & 3137 & $16.3302\pm0.0019$ & 1318 & $1.1254\pm0.0031$ & 389 & $1.0226\pm0.0128$ & 33 &      &     & $a$ \\
     & 5719 & 3523 & $16.3209\pm0.0017$ & 1434 & $1.0906\pm0.0029$ & 680 & $1.0328\pm0.0165$ & 35 &      &     & \\
     & 5430 & 3014 & $16.3714\pm0.0017$ & 1626 & $1.1235\pm0.0027$ & 735 & $1.0223\pm0.0086$ & 34 &      &     & \\
     & 5746 & 3556 & $16.3931\pm0.0019$ & 1723 & $1.1230\pm0.0025$ & 745 & $1.0665\pm0.0141$ & 33 &      &     & $a$ \\
     & 6038 & 4036 & $16.4230\pm0.0020$ & 1320 & $1.1336\pm0.0028$ & 517 & $1.0592\pm0.0099$ & 33 &      &     & $a$ \\
     & 5804 & 3655 & $16.5571\pm0.0020$ & 1626 & $1.1636\pm0.0030$ & 457 & $1.0947\pm0.0232$ & 31 &      &     & $a$ \\
     & 5500 & 3139 & $16.5705\pm0.0014$ & 2001 & $1.1799\pm0.0024$ & 593 & $1.1063\pm0.0159$ & 33 &      &     & \\
     & 6043 & 4048 & $16.5880\pm0.0028$ & 1486 & $1.1862\pm0.0036$ & 411 & $1.0765\pm0.0405$ & 25 &      &     & \\
     & 5363 & 2877 & $16.7291\pm0.0017$ & 1532 & $1.2342\pm0.0025$ & 719 & $1.1475\pm0.0213$ & 31 &      &     & \\
     & 5710 & 3508 & $16.6680\pm0.0021$ & 1378 & $1.2011\pm0.0030$ & 658 & $1.1197\pm0.0130$ & 32 &      &     & \\
     & 5839 & 3704 & $16.6637\pm0.0020$ & 2015 & $1.1962\pm0.0029$ & 585 & $1.0950\pm0.0156$ & 33 &      &     & $a$ \\
     & 5390 & 2923 & $16.7092\pm0.0018$ & 1900 & $1.2129\pm0.0025$ & 584 & $1.1296\pm0.0134$ & 31 &      &     & $a$ \\
     & 5549 & 3215 & $16.8581\pm0.0204$ &   72 & $1.2498\pm0.0239$ &  11 &                 &  0 &      &     & $b$ \\
     & 5700 & 3487 & $16.9410\pm0.0045$ & 1246 & $1.2737\pm0.0052$ & 646 & $1.1191\pm0.0256$ & 32 &      &     & $a$ \\
     & 5482 & 3103 & $17.0145\pm0.0018$ & 1843 & $1.3203\pm0.0029$ & 568 & $1.2082\pm0.0301$ & 34 &      &     & \\
     & 6152 & 4297 & $17.1411\pm0.0036$ &  236 & $1.3947\pm0.0069$ & 146 &                 &  0 &     &     & \\
     & 5355 & 2857 & $17.1561\pm0.0022$ & 1538 & $1.3598\pm0.0035$ & 565 & $1.2381\pm0.0206$ & 29 &     &     & $a$ \\
     & 5397 & 2943 & $17.1754\pm0.0020$ & 1672 & $1.4189\pm0.0034$ & 670 & $1.2812\pm0.0176$ & 29 &     &     & \\
     & 6044 & 4035 & $17.2250\pm0.0032$ & 1284 & $1.3921\pm0.0043$ & 554 & $1.2759\pm0.0234$ & 31 &     &     & $a$ \\
     & 5275 & 2700 & $17.2840\pm0.0036$ &  213 & $1.4352\pm0.0046$ & 394 & $1.2553\pm0.0433$ & 26 &     &     & \\
     & 5919 & 3824 & $17.3165\pm0.0033$ & 1607 & $1.4209\pm0.0055$ & 285 & $1.2715\pm0.0424$ & 22 &     &     & \\
     & 6100 & 4130 & $17.3778\pm0.0044$ &  204 & $1.4360\pm0.0063$ & 274 & $1.2827\pm0.0468$ & 26 &     &     & \\
     & 5561 & 3262 & $17.4041\pm0.0029$ &  101 & $1.4551\pm0.0038$ & 678 & $1.3116\pm0.0210$ & 24 &     &     & \\
     & 5244 & 2601 & $17.4292\pm0.0038$ &   77 & $1.4761\pm0.0102$ &  75 & $1.3315\pm0.0460$ & 17 &     &     & \\
     & 5720 & 3534 & $17.4359\pm0.0102$ & 1343 & $1.4612\pm0.0138$ & 169 &                 &  0 &     &     & \\
     & 5891 & 3801 & $17.5168\pm0.0030$ & 1539 & $1.4632\pm0.0043$ & 605 & $1.3240\pm0.0361$ & 21 &     &     & \\
     & 5399 & 2948 & $17.5722\pm0.0035$ &  193 & $1.5123\pm0.0044$ & 536 & $1.3234\pm0.0226$ & 24 &     &     & \\
     & 5856 & 3727 & $17.6321\pm0.0032$ &   75 & $1.5514\pm0.0045$ & 675 & $1.4138\pm0.0401$ & 19 &     &     & \\
     & 5480 & 3100 & $17.6454\pm0.0035$ & 1365 & $1.5545\pm0.0049$ & 538 & $1.3281\pm0.0358$ & 25 &     &     & \\
     &      & 3759 & $17.6587\pm0.0034$ &   42 & $1.5293\pm0.0040$ & 652 & $1.3741\pm0.0526$ & 19 &     &     & \\
     & 5875 & 3774 & $17.6647\pm0.0058$ &  353 & $1.5763\pm0.0081$ & 642 & $1.3384\pm0.0408$ & 18 &     &     & \\
\end{tabular}
\end{minipage}
\end{table*}

\begin{table*}
\begin{minipage}{15cm}
\contcaption{}
\begin{tabular}{@{}ccccrcrcrrrl@{}}
\hline
Sanders & MMJ & Fan & $V$ & $N_{V}$ & $V-I$ & $N_{I}$ & $B-V$ & 
$N_{B}$ & $P_{S}$ & $P_{G}$ & Notes \\
\hline
     & 5281 & 2713 & $17.7251\pm0.0065$ &   73 & $1.5160\pm0.0075$ & 273 &                 &  0 &     &     & $b$ \\
     & 5861 & 3748 & $17.7940\pm0.0038$ & 1172 & $1.5783\pm0.0044$ & 648 & $1.3792\pm0.0556$ & 16 &     &     & \\
     & 5916 & 3845 & $17.7913\pm0.0035$ &  124 & $1.5797\pm0.0044$ & 585 & $1.4139\pm0.0415$ & 16 &     &     & \\
     &      & 3636 & $17.8381\pm0.0060$ &  396 & $1.5984\pm0.0072$ & 631 & $1.3349\pm0.0926$ & 13 &     &     & \\
\multicolumn{12}{c}{\underline{Beyond Color Range of Calibration Stars}} \\
     & 5802 & 3620 & 17.95              &  806 & 1.60              & 162 &                 &  0 &     &     & \\
     &      & 3322 & 17.97              &   59 & 1.64              & 625 & 1.40              & 14 &     &     & \\
     & 5475 & 3090 & 18.00              & 1027 & 1.62              & 625 & 1.44              & 14 &     &     & \\
     & 5255 & 2650 & 18.04              &  112 & 1.66              & 321 & 1.44              & 17 &     &     & \\
     & 5456 & 3061 & 18.11              &  522 & 1.68              & 607 & 1.47              & 11 &     &     & \\
     & 5957 & 3891 & 18.12              &  187 & 1.67              & 583 &                 &  0 &     &     & \\
     & 5811 & 3666 & 18.16              &   57 & 1.73              & 364 &                 &  0 &     &     & \\
     & 5235 & 2590 & 18.33              &   36 & 1.75              & 279 &                 &  0 &     &     & \\
     & 6115 & 4181 & 18.33              &  405 & 1.77              & 229 &                 &  0 &     &     & \\
     & 6170 & 4331 & 18.43              &   18 & 1.81              &  22 &                 &  0 &     &     & \\
     & 5550 & 3244 & 18.48              &  312 & 1.79              & 552 &                 &  0 &     &     & \\
     & 5231 & 2589 & 18.53              &  320 & 1.87              & 270 &                 &  0 &     &     & \\
     &      & 4009 & 18.57              &  264 & 1.84              & 453 &                 &  0 &     &     & \\
     & 5202 & 2527 & 18.57              &   93 & 1.83              & 154 &                 &  0 &     &     & \\
     & 6207 & 4402 & 18.58              &   36 & 1.82              &  80 &                 &  0 &     &     & \\
     & 5291 & 2731 & 18.62              &   66 & 1.90              & 385 &                 &  0 &     &     & \\
     & 5400 & 2949 & 18.63              &  347 & 1.80              & 511 &                 &  0 &     &     & \\
     & 6150 & 4276 & 18.67              &  178 & 1.90              & 160 &                 &  0 &     &     & \\
     & 5650 & 3415 & 18.79              &   82 & 1.88              & 517 &                 &  0 &     &     & \\
\hline
\end{tabular}

\medskip
Notes: $a$: Star used in the ``bluest star'' sequence.

$b$: Star not used in the determining fiducial line.

BSS?: possible blue straggler star
\end{minipage}
\end{table*}

In addition we identify an alternate sequence of main sequence stars
that fall at the blue end of the color distribution at each magnitude
level.  If our magnitudes are mostly free of systematic errors and our
error estimates reflect the real uncertainty in measurements relative
to other stars, then these stars would most accurately delineate the
SSS. The stars in this sequence are notated with a ``$b$'' in Table
\ref{sss}. In most cases, the difference in color between this bluer
sequence and the average sequence is no more than 0.02 in $V-I$.

Finally, in Table \ref{weird}, we identify stars that are worthy of
further study. We have identified several stars on the RGB that are
likely to be part of binary systems, several new blue straggler star
(BSS) candidates, and possible triple systems.

\begin{table*}
\begin{minipage}{170mm}
\caption{Unusual Stars in M67}
\label{weird}
\begin{tabular}{@{}ccccrcrcrrrl@{}}
\hline
Sanders & MMJ & Fan & $V$ & $N_{V}$ & $V-I$ & $N_{I}$ & $B-V$ & $N_{B}$ &
$P_{S}$ & $P_{G}$ & Notes \\
\hline
1054 & 6489 & 3347 & $11.1393\pm0.0005$ & 2239 & $1.0706\pm0.0009$ & 423 & $1.0810\pm0.0024$ & 62 & 0.64 & 0.99 & RGB binary?\\
1288 & 6505 & 4118 & $11.2502\pm0.0005$ & 1606 & $1.0560\pm0.0014$ & 292 & $1.0714\pm0.0015$ & 62 & 0.96 & 0.99 & RGB binary?\\
1254 & 6500 & 4187 & $11.4899\pm0.0005$ &  883 & $1.0431\pm0.0007$ & 296 & $1.0323\pm0.0037$ & 56 & 0.95 & 0.99 & RGB binary?\\
1293 & 6050 & 4039 & $12.1223\pm0.0004$ & 2163 & $0.9936\pm0.0006$ & 636 & $0.9952\pm0.0017$ & 62 & 0.93 & 0.97 & RGB binary?\\
1001 &      & 3075 & $12.3926\pm0.0005$ & 1926 & $0.9874\pm0.0006$ & 819 & $0.9659\pm0.0056$ & 62 & 0.95 & 0.99 & RGB binary?\\
489  &      & 1770 & 12.76              &      &                   &     & 0.68              &    & 0.95 & 0.97 & TO gap \\
615  & 5118 & 2314 & 12.818             &      & 0.653             &     & 0.521             &    & 0.96 & 0.98 & TO gap \\
1271 & 5969 & 3904 & $12.8502\pm0.0006$ & 1400 & $0.6148\pm0.0007$ & 772 & $0.5238\pm0.0016$ & 62 & 0.94 & 0.99 & TO gap \\
794  & 5318 & 2794 & $12.8650\pm0.0004$ & 2135 & $0.9674\pm0.0009$ & 563 & $0.9478\pm0.0048$ & 37 & 0.93 & 0.98 & RGB binary?\\
610  &      & 2164 & 12.91              &      &                   &     & 0.49              &    & 0.88 & 0.98 & TO gap \\
1463 & 6259 & 4505 & $12.920\pm0.015$   &      & $1.014\pm0.011$   &     & $0.983\pm0.023$   &    & 0.96 & 0.98 & RGB triple? \\
602  &      & 2026 & 12.94              &      &                   &     & 0.54              &    & 0.93 & 0.99 & TO gap \\
1503 &      & 4705 & 13.05              &      &                   &     & 0.54              &    & 0.87 & 0.93 & TO gap \\
1575 &      & 4970 &                    &      &                   &     &                   &    & 0.91 & 0.98 & TO gap \\
2219 & 5603 & 3345 & $13.1829\pm0.0004$ & 2036 & $0.6227\pm0.0006$ & 814 & $0.5534\pm0.0033$ & 62 & 0.99 & 0.99 & BSS? \\
1292 & 5961 & 3889 & $13.1974\pm0.0003$ & 2434 & $0.6936\pm0.0006$ & 774 & $0.6178\pm0.0027$ & 62 & 0.95 & 0.98 & triple? \\
816  &      & 2829 & $13.2527\pm0.0013$ &  411 & $0.6938\pm0.0042$ &  26 & $0.8409\pm0.0207$ & 13 & 0.95 & 0.98 & triple? \\ 
1220 & 6125 & 4214 & $13.3164\pm0.0053$ &  354 & $0.6211\pm0.0055$ & 272 & $0.5806\pm0.0062$ & 51 & 0.95 & 0.99 & BSS? \\
773  & 5377 & 2904 & $13.3214\pm0.0013$ & 1876 & $0.7120\pm0.0014$ & 812 & $0.5826\pm0.0041$ & 62 & 0.96 & 0.99 & triple? \\
1226 & 6112 & 4179 & $13.3357\pm0.0010$ &  989 & $0.6336\pm0.0013$ & 356 & $0.5499\pm0.0036$ & 55 & 0.92 & 0.98 & BSS? \\
856  &      & 2734 &                    &      &                   &     &                   &    & 0.90 & 0.87 & BSS? \\
927  &      & 3239 &                    &      &                   &     &                   &    & 0.96 & 0.98 & BSS? \\
1011 & 5844 & 3717 & $13.8050\pm0.0006$ & 1996 & $0.7548\pm0.0008$ & 811 & $0.6306\pm0.0023$ & 61 & 0.93 & 0.97 & triple? \\
1608 & 6443 & 4956 & $14.248\pm0.007$   &      & $0.833\pm0.003$   &     & $0.775\pm0.011$   &    & 0.72 & 0.86 & triple? \\
1249 & 6144 & 4258 & $14.3042\pm0.0013$ &  663 & $0.8647\pm0.0022$ & 274 & $0.7813\pm0.0877$ & 30 & 0.92 & 0.92 & triple?\\
1492 &      & 4654 & 14.56              &      &                   &     & 0.85              &    & 0.85 & 0.01 & triple? \\
787  & 5387 & 2922 & $14.5636\pm0.0008$ & 2280 & $0.9277\pm0.0011$ & 755 & $0.8610\pm0.0035$ & 59 & 0.05 & 0.43 & member?\\
1601 &      & 5018 & 14.57              &      &                   &     & 1.02              &    & 0.91 & 0.48 & triple? \\
475  &      & 1826 &                    &      &                   &     &                   &    & 0.92 & 0.00 & triple? \\
1209 & 6041 & 4047 & $15.6339\pm0.0020$ & 1611 & $0.6141\pm0.0026$ & 359 & $0.4907\pm0.0255$ & 28 & 0.96 &      & ?? \\
\hline
\end{tabular}
\end{minipage}
\end{table*}

\subsubsection{The Upper Red Giant Branch (RGB)}

The observations made by us and by Fan et al. were generally long
enough that stars on the upper giant branch were saturated on the CCD,
and so were not measured by either study. However, Janes \& Smith
(1984) studied the upper giant branch to try and sort
first-ascent giant stars from stars in later evolutionary phases and
star in binaries. \citet{mmj} tabulate photometry for the brightest
stars taken from other studies. Because there are likely to be
systematic errors resulting from differences in calibration, we do not
include these stars in Table \ref{sss}. However, the stars that are
most likely to be on the SSS (in order from RGB tip faintward) are
S1553 (61\%), S488 (51\%), S1135 (19\%), S978 (99\%; 95\%), S364
(82\%), S1557 (77\%), and S1016 (98\%; 93\%). The percentages quoted
are proper motion membership probabilities from \citet{sanders} and
\citet{girard} if available. All of these stars are high probability
proper motion and radial velocity members \citep{mathieurv} with the
exception of S1135, which is a high-probability radial velocity
member.

\subsubsection{The Red Giant Clump}

We were able to photometer 4 stars falling in the red giant clump (see
Table \ref{sss}), and identify 3 others (S1316, S1479, and S1592) from
the \citet{mmj} listing of bright stars. All are high probability
proper motion and radial velocity members of M67.

\subsubsection{The Lower Giant Branch\label{lrgb}}

\begin{figure*}
\includegraphics[width=160mm]{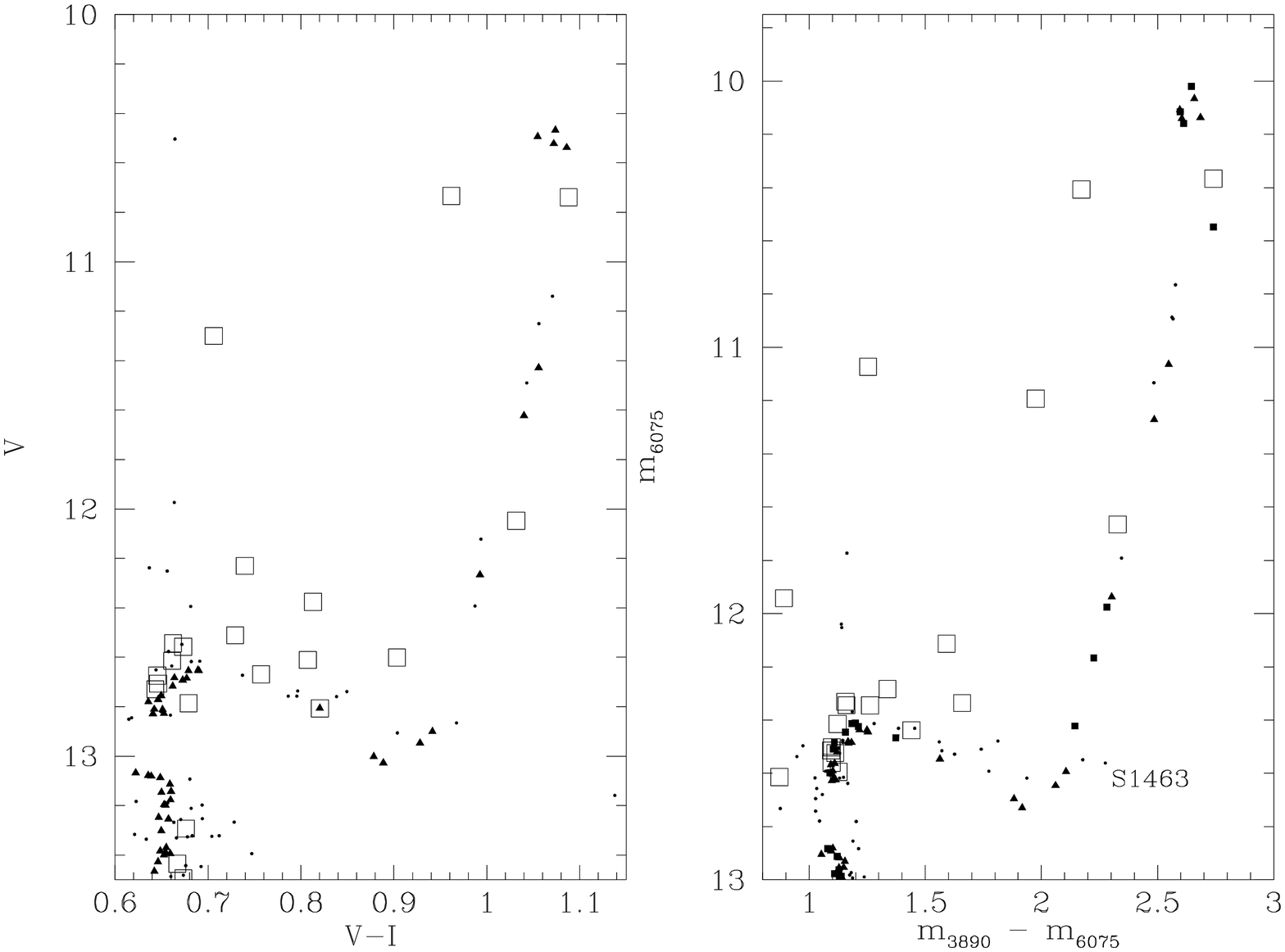}
\caption{CMDs covering the red giant branch (RGB) and subgiant branch
(SGB) from this study ({\it left panel}) and Fan et al. (1996; {\it
right panel}), including stars selected for the $VI$ fiducial line
({\it triangles}), additional stars on the fiducial that were not
observed in this study ({\it filled squares}), and known binaries
({\it open squares}).
\label{rgbsgb}}
\end{figure*}

We selected giant branch members based on whether they seemed to be
closest to a lower envelope of stars leading up the RGB.  In many
cases, cluster members fall near, but not on the giant branch.  While
some of these are to the blue of the giant branch (and thus are very
likely to be binary stars), there are several stars to the {\it red}
of the giant branch. Among the bluer stars are known binaries like
S1040, S1182, S1195, S1221, and S1237. We also identify S721, S1054,
S1254, and S1288 as probable binaries based on their colors from
our data and \citet{fan} data. We also reject S1293 as too blue in our
data, although it appears to lie on the giant branch in the Fan et
al. data. CMDs are shown in Fig. \ref{rgbsgb}.

As mentioned earlier, systems can be created with colors {\it redward}
of the giant branch if a main sequence secondary star with a color
redder than the giant star is present. S1001 is one example that
appears redward of the giant branch by a small but noticeable amount
in all of the colors we examined, so that it was not included in the SSS.
The fainter the giant star is, the larger the red deviation caused by
a main sequence companion can be. S794 is an example of this near the
base of the red giant branch. S1463 is another high probability
cluster member nearby in the CMD that deviates
by an amount that {\it cannot} be explained by a single main sequence
companion.

One can also question whether surface composition changes driven by
evolutionary changes at first dredge-up on the RGB might significantly
affect the CMD position of stars in this section. This is only likely
to be a significant worry if different M67 giants undergo the
dredge-up event at substantially different positions on the RGB or if
there is scatter in the amount of material dredged up. Brown (1987)
examined C and N abundances along the giant branch, and found a fairly
abrupt change in $12.2 < V < 13$. Gilroy \& Brown (1991) examined
$^{12}$C / $^{13}$C ratios, and found them varying in the same
range. This is partly because there are few stars in this portion of
the CMD (see Fig. \ref{rgbsgb}). In fact, we rejected only three stars
in this range: S794, S1001, and S1463. S794
was observed by Gilroy \& Brown though, and seems to have abundances
consistent with not yet having undergone dredge-up. S1463 was also
observed, but its values were inconclusive. However, Balachandran
(1995) detected Li in the star, which is unusual because none of the
other stars in similar evolutionary stages have detectable Li. We
believe this spectroscopic evidence strongly supports the exclusion of
S794 and S1463 from the SSS. S1001 remains a marginal case.

\subsubsection{The Subgiant Branch (SGB)}

We define this region to be between the base of RGB (the local minimum
in luminosity) and the top of the main sequence (the local maximum).
A close examination of the ($m_{3890}, m_{3890}-m_{9190}$) CMD from
\citet{fan} reveals a well-defined string of stars delineating the
SGB. In the ($m_{6075}, m_{3890}-m_{6075}$) CMD, there are clearly a
small number of stars that are fainter than the rest. There is one
known binary (S1242) that could be mistaken for a normal SGB star in
the ($m_{3890}, m_{3890}-m_{9190}$) CMD, but falls among the more
numerous and brighter subset of stars in ($m_{6075},
m_{3890}-m_{6075}$) CMD.  Since blends {\it only} fall above the SGB
in this portion of either CMD, we select the fainter subset as the
likely SSS. The neater appearance of the SGB in ($m_{3890},
m_{3890}-m_{9190}$) is due to the large wavelength separation of the
passbands used in the color, which can hide small deviations from the
SSS. This should serve as a reminder of the importance of multi-band
imaging.  The difference between the bright and faint subsets is a few
hundredths of a magnitude in $m_{6075}$.

An implication of this is that the majority of the stars seen near the
cluster SGB may be binary stars: S591, S781, S961, S1069, S1239, S1438,
and S1487. We have looked at the radial velocities tabulated by
\citet{mathieurv} for most of our selected stars and these presumed
binaries.  There were three stars originally in our presumed single
group (S1060, S1323, and S2207) and two stars (S961 and S1239) having
radial velocity dispersions labelled ``hi'' ($\sigma \ge 1.0$
\kms) by \citet{girard}. However, only S2207 has a large number of
measurements and $\sigma > 2$ \kms. We have opted to retain the other
two ``single'' stars in our sample to flesh out the SGB, but this fact should 
be kept in mind.

At the blue end of the SGB (the local maximum in luminosity of the SSS), it
becomes impossible to unambiguously distinguish binaries from single
stars based on photometry alone because blends with faint main sequence
stars move the system parallel to the SSS. There are a number of stars
in this portion of the diagram that could be single stars, so we resort to
simply tabulating all of these and averaging the photometry so that we
are not too far off in magnitude. 

As for strange stars, Mathieu et al. (2003) have recently discussed
the two ``sub-subgiant'' branch stars S1063 and S1113. Both are known
binaries with period of several days, and both are high-probability
proper motion members. There is not currently an accepted explanation
for the positions of these two stars in the CMD.

\subsubsection{The Upper Turnoff}

\begin{figure*}
\includegraphics[width=160mm]{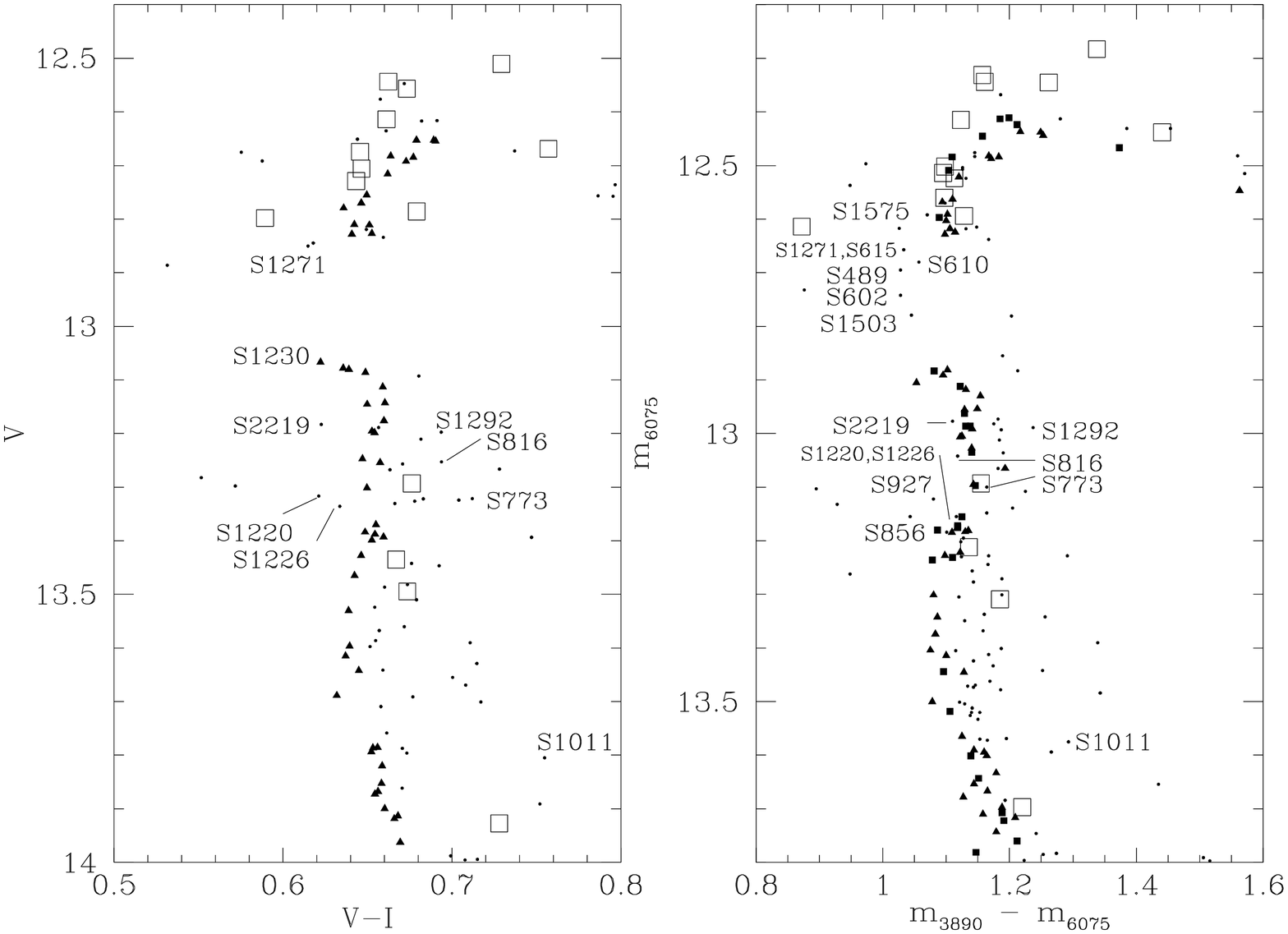}
\caption{CMDs covering the cluster turnoff. Symbols have the same
meaning as in Fig. \ref{rgbsgb}.
\label{cmdto}}
\end{figure*}

We define this region to be the portion of the SSS between the gap
and the local luminosity maximum ($12.65 < V < 12.85$). The number of
known binaries falling in this portion of the diagram again indicates
that contamination by unresolved blends of stars may be a serious problem.
Fig. \ref{cmdto} shows an exploded view of this part of the CMD.

\begin{figure*}
\includegraphics[width=160mm]{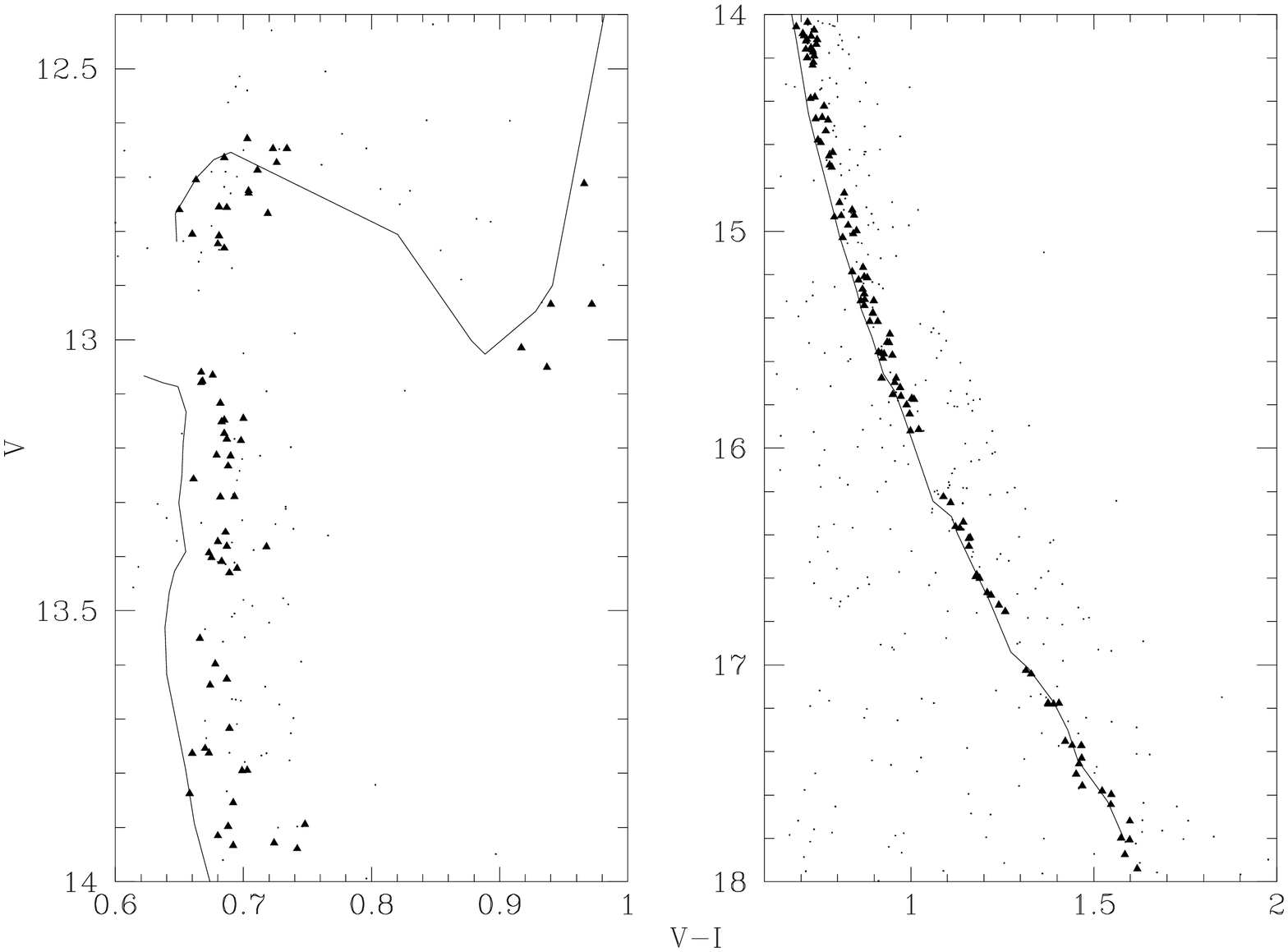}
\caption{$VI$ CMD for M67 using data from \citet{mmj}. The selected
single stars from Table \ref{sss} are plotted as triangles. The solid lines 
are fiducial points from this study (Table \ref{fiduc}).
\label{mmjcmd}}
\end{figure*}

We can get a degree of reassurance from the fact that a number of
these systems (specifically S926, S995, S1310, S1589, and S1639) do
have multiple radial velocity observations that are consistent both
with the cluster mean and with no variation. Once again, we decided to be
conservative in our selection in this region, retaining stars and averaging
their photometry so that the fiducial could not be far off the SSS.
For comparison,
we plot corresponding regions of the ($V,V-I$) CMD for the 
\citet{mmj} dataset in Fig. \ref{mmjcmd}. This figure shows the improvement
in the photometric scatter near the turnoff, and illustrates the 
systematic differences stemming from the calibration to the standard system.

\subsubsection{The Turnoff Gap}

Many previous investigators have drawn attention to the gap at the
turnoff between $12.85 < V < 13.05$. The gap is the result
of short evolutionary time-scales for single stars in this portion of
the diagram. We will discuss the phase of single star evolution that
is likely to produce this gap in \S \ref{toshape} below. However, we
note that there are a number of proper motion members that do fall in
this range of magnitudes. There is a possibility that these stars
could be single stars, although we believe it is more likely that they
should be classified as blue stragglers. Only one cluster member (S1271)
fell in the fields we observed, but an examination of the Fan et
al. data reveals an additional six (S489, S602, S610, S615, S1503,
S1575), none of which has published radial velocity observations.

\subsubsection{The Lower Turnoff\label{lto}}

For the lower portion of the turnoff region ($13.05 < V < 13.6$), it is
once again possible to start to apply a ``bluest star'' criterion for
determining the position of the SSS since the blend sequences no
longer intersect the SSS except near the very reddest portion. We
still need to be aware of the possibility that some blue stragglers
may have to be eliminated. Unfortunately the faint limit of the
\citet{mathieurv} radial velocity study falls in this range of
magnitudes also, so that binaries can now only be eliminated based on
photometric variations or on CMD position.

The first apparent feature is formed by four stars (S731, S998, S1230,
and S1302) that form a nearly linear feature in all of the CMDs we
have examined. All four are high probability cluster members and show
no signs of photometric variability (this work; S998, S1230, S1302 in
Stassun et al. 2002; S998 in Gilliland et al. 1991), but unfortunately
only S998 has been observed spectroscopically. S998 was observed by
\citet{melo}, who found a radial velocity consistent with the cluster
mean, and a low rotational velocity ($v \sin i = 6.3 \pm 0.6$ \kms)
consistent with those observed for other turnoff stars. Li has also
been detected at the surface of S998 by \citet{hobbs} at a level
consistent with other stars at the turnoff. (To date, there have not
been any confirmed detections of lithium in M67 blue stragglers.)
While there is no reason to doubt cluster membership and no evidence
of variability or of companion stars, the small number of stars
involved means that this should tested further. For example, the
bluest star S1230 can be questioned based on its slight fainter
$m_{6075}$ magnitude in the Fan et al. CMD. We will discuss these
stars more in \S \ref{toshape}.

While our color resolution of the turnoff region is somewhat limited,
there is clear evidence that stars in this range of magnitudes have
evolved to cooler surface temperatures. In this range of magnitudes we
begin to tabulate two separate sequences: one based on the bluest
reliable cluster member, and one based on weighted averages of groups
of stars near the blue edge of the star distribution. In both cases a
redward turn is noticeable.  Several stars in our $VI$ CMD can be
tentatively classified as blue stragglers based on their CMD positions
bluer than the faint portion of the turnoff and fainter than the
bright portion of the hook. S856 and S927 were only observed by
\citet{fan}, but have high membership probabilities. S1220, S1226, and S2219 were observed
by us and all have high membership probabilities (see Table
\ref{weird}). However, in the Fan et al. dataset, these stars fall
closer to the presumed SSS, although on the blue side of the
distribution of colors. The weakest candidate is S1220, which had
the largest scatter in measurements and the fewest measurements
overall. Other blue stragglers have been discussed elsewhere (e.g.,
Sandquist \& Shetrone 2003b)

Finally, we have looked for cluster members that fall to the red of
the two-star blend sequence since stars might be present in this
portion of the CMD if they are part of a triple system. We find 4
candidates in our photometry (S773, S816, S1011, and S1292), although
these fall within the two-star blend sequence in the Fan et
al. ($m_{6075}, m_{3860} - m_{6075}$) CMD. This may be the result of much
lower surface temperature for the companion star or stars, since the
color in the Fan et al. data more closely resembles $B-V$. We call
attention to these stars to encourage further examination.

\subsubsection{The Upper Main Sequence}

As mentioned in the previous subsection, the SSS for the main sequence
can be most cleanly identified using the best measured blue stars at a
given magnitude level. We can use proper motion information on the
upper main sequence to eliminate potential confusion from the small
numbers of field stars that fall very close to the main sequence in
the CMD. However, as we have also mentioned, it is apparent that the
proper motion membership information becomes less trustworthy for the
faintest stars in the samples of \citet{sanders} and
\citet{girard}. 

For $V \ga 15.2$, we allow stars with low proper motion
membership probabilities into the sample if they clearly fell near
other proper motion members on the main sequence.  The faintest star
selected for the SSS that has high membership probabilities according
to both \citet{sanders} and \citet{girard} is S1291 at $V = 15.29$, so
that fainter stars that we retained had low membership probabilities
according to either \citet{sanders} or \citet{girard}. The faintest
star with high probability in just one of the two studies is S769 at $V =
15.61$. At this point the proper motion surveys begin to be
seriously incomplete. We do not believe these difficulties bias our
measured position for the SSS because the main sequence
can be clearly traced throughout.

The star S1209 has a 96\% membership probability according to
\citet{sanders}, but is much too blue to be a main sequence star or a
blend with a white dwarf star. If this star is indeed a cluster
member, we do not have a good explanation for its photometry. We also
note that there are several additional stars that fall significantly
to the red of the blend sequence in this portion of the color
magnitude diagram: S787 and S1249 according to our photometry, and
S475, S1492, S1601, and S1608 from \citet{fan}. Unfortunately, the
proper motion information disagrees on membership for all of these
stars except S1249 and S1608. We note that the binary star S1019
\citep{vSVM} also falls in this portion of the diagram. S1019 is a
known X-ray source containing a short-period binary, which makes it
possible that another star is present on a much larger orbit. Other
systems found to the red of the blend sequence have proper motions
that indicate they are nonmembers.

\subsubsection{The Lower Main Sequence}

\begin{figure*}
\includegraphics[width=160mm]{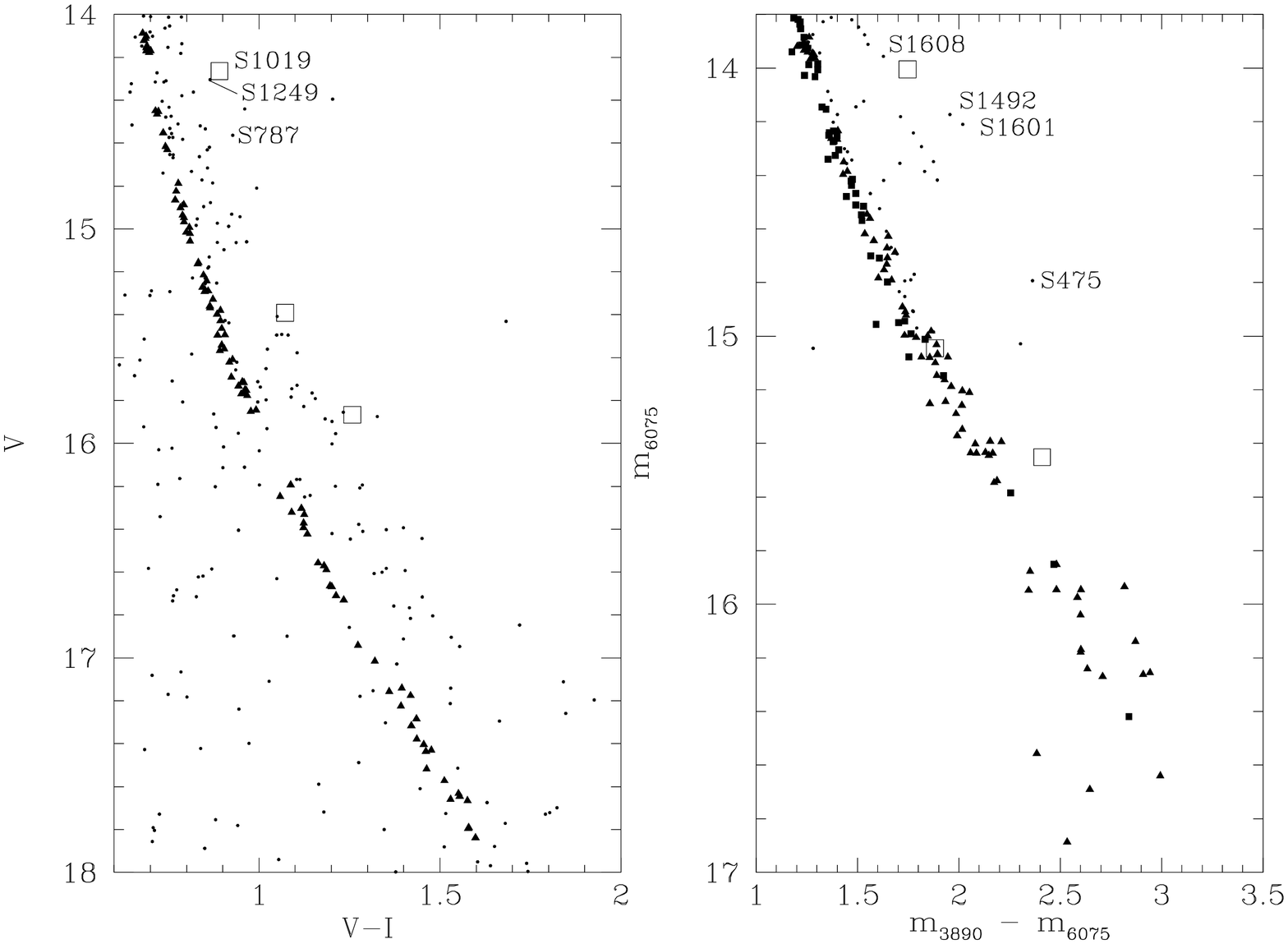}
\caption{CMDs covering the cluster main sequence. Symbols have the same
meaning as in Fig. \ref{rgbsgb}. Dots in the
right panel do not represent all of the stars observed by Fan et al. (1996):
only those with proper motion information. The faint limit
of the $m_{3890}$ data falls within the diagram.
\label{cmdms}}
\end{figure*}

For the faintest portions of the main sequence that we studied, CMD
position is the only criterion we can use to select stars. Another
well-known gap is present on the main sequence at $15.9 \la V \la
16.2$, although there is no obvious explanation for this. As can be
seen from Fig. \ref{cmdms}, the Fan et al. dataset is no longer useful
for selecting stars for the SSS below the gap due to large photometric
scatter as the faint limit of the 3890 \AA observations is approached
at $m_{3890} \approx 20$. We terminated our tabulation of the fiducial
line at $V \approx 17.9$, because the color terms in the
photometric transformations cannot be reliably extrapolated beyond the
end of the color range of our calibrating stars $(V-I) \approx
1.6$. We have, however, tabulated approximate magnitudes and colors
for fainter stars that fall on an extension of the main sequence.

\section{Discussion}

\subsection{Distance Modulus}

Percival, Salaris, \& Kilkenny (2003) have recently published a sample
of local G and K dwarfs with accurate Hipparcos parallaxes in a
metallicity range appropriate for open cluster main sequence fitting.
Percival et al. also describe a method of correcting the dwarf colors
for metallicity differences with the cluster that does not require
theoretical isochrones.  With our new photometric calibration and
determination of the cluster's fiducial line, it is worthwhile to
attempt to update the distance modulus for the cluster.

\begin{figure}
\includegraphics[width=84mm]{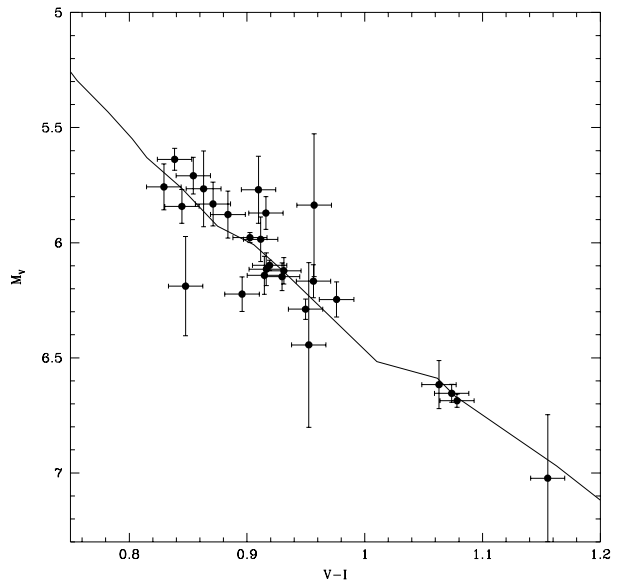}
\caption{The best fit of our $VI$ fiducial line to the metal-rich dwarf star
sample of Percival et al. (2003), corresponding to $(m - M)_0 = 9.60$.
\label{dm}}
\end{figure}

In doing the fits to our fiducial points, we eliminated local dwarfs
having $(V-I) < 0.83$ from the sample in order to remove any possible
effects that stellar evolution might have. [The choice of color cutoff
was based on the isochrones of Girardi et al. (2000).] A total of 27
dwarfs with $(V-I) < 1.16$ were used in the fit. We assumed [Fe/H] $=
0.02 \pm 0.06$ (Gratton 2000) and E$(B-V) = 0.04 \pm 0.01$
(corresponding to E$(V-I) = 0.050 \pm 0.013$). We find $(m - M)_0 =
9.60 \pm 0.03$ [$(m - M)_V = 9.72 \pm 0.05$], where the uncertainty
includes uncertainty in the fit and in the reddening. The best fit to
our fiducial line is shown in Fig. \ref{dm}. Our derived distance
modulus is significantly lower (0.12 mag) than the value given by
\citet{mmj} for an isochrone fit to their $VI$ dataset ($(m - M)_V =
9.85$, or $(m-M)_0 = 9.72$), but in good agreement with a recent
isochrone fit to the $BV$ data of \cite{mmj} by Sarajedini et
al. (1999), giving $(m-M)_V = 9.69 \pm 0.11$. The difference in the
distance moduli derived from the $VI$ datasets stems from the
calibration to the standard system, as can be seen in
Fig. \ref{mmjcmd}.

\subsection{Comparison with Theoretical Isochrones}

A careful comparison between our best measured highest-probability
single stars and theoretical isochrones reveals a number of areas of
significant disagreement. We focus on recent sets of models by Girardi
et al. (2000; hereafter, referred to as Padova isochrones), Yi et
al. (2001; hereafter, the Yonsei-Yale or Y$^2$ isochrones --- we use
version 2 of these isochrones), and Baraffe et al. (1998; hereafter
BCAH).

\subsubsection{Absolute and Relative Colors of the Upper Main Sequence and Lower Giant Branch\label{msrgb}}

Predicted colors depend critically on the choice of
$T_{\mbox{eff}}$-to-color transformation.  The color transformations
used in the Padova isochrones are described in Girardi et
al. (2002). For most of the area of the CMD we consider, the
theoretical colors are based on ATLAS9 synthetic stellar atmospheres
(Kurucz 1993) recalculated by Castelli, Gratton, \& Kurucz (1997). The
Y$^2$ isochrones employ transformation tables from two different
sources: Lejeune, Cuisinier, \& Buser (1998; hereafter LCB) and Green,
Demarque, \& King al. (1987; hereafter GDK). GDK made use of older
Kurucz (1979) atmosphere models to predict colors. LCB used empirical
solar-abundance color-temperature relations to make corrections to
their library spectra (Kurucz 1993) so that the spectra produce
color-temperature relations in agreement with empirical
solar-abundance relations. Corrected theoretical spectra can then be
used to extended the relations to other abundances in a differential
sense (Lejeune, Cuisinier, \& Buser 1997). Their computed
solar-abundance relations generally agree with the empirical ones to
better than 0.1 mag. Since M67 is very close to solar abundance, this
means that the use of LCB tables should almost be equivalent to using
empirical color-temperature relations.

In a direct comparison between the models and observations, there are
differences in the color positions of the unevolved model main
sequences. Some error in absolute color is to
be expected because the line lists and line parameters used in the
theoretical atmosphere models do not accurately reproduce
high-resolution spectra (see, for example, the discussion in \S 3.1.2
of \citet{gir02}), but generally the hope is that these errors
average out, particularly when wide bandpasses are used. However,
the Padova isochrones are noticeably redder than the Y$^2$
isochrones with LCB tables at all magnitude levels, in spite of the
use of the same set of stellar atmospheres. The smallest color
difference occurs near where the solar calibration requires better
agreement. Thus, the differences in absolute colors seem to be the
result of the {\it corrections} made to the synthetic photometry.

\begin{figure*}
\includegraphics[width=160mm]{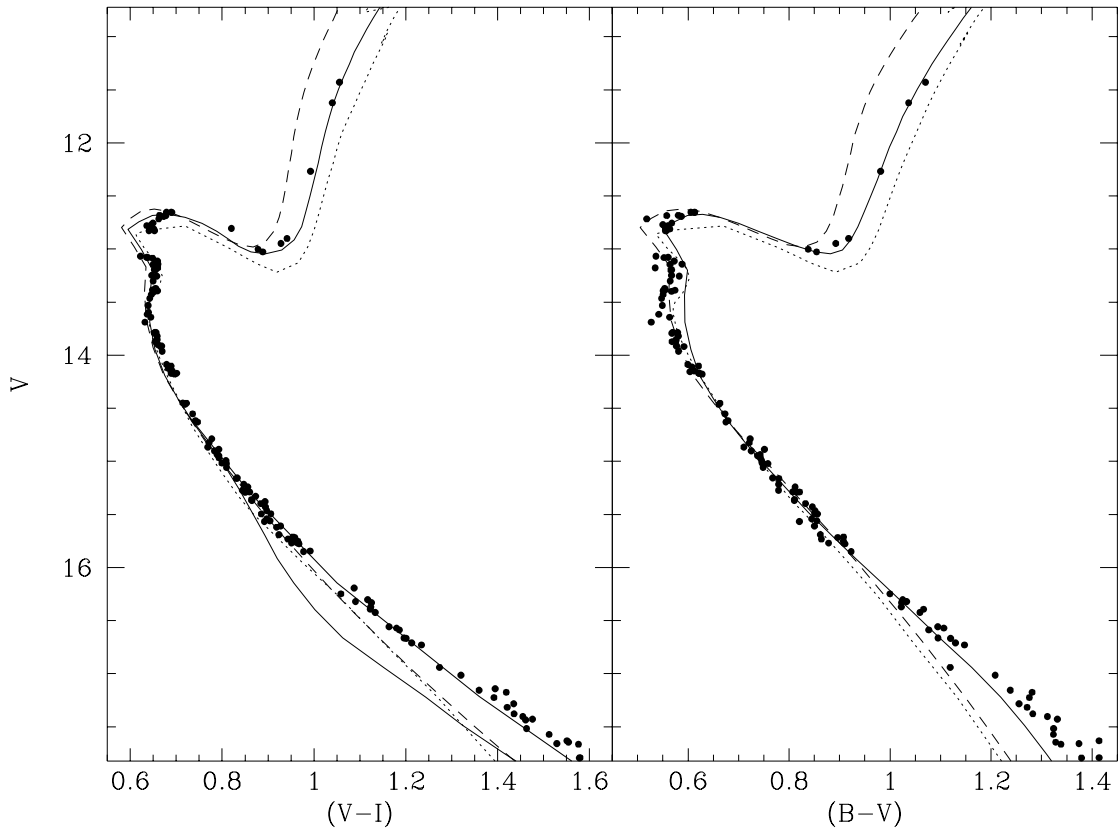}
\caption{A comparison of our selected M67 stars ({\it solid points})
with theoretical isochrones for 4 Gyr from the Yale-Yonsei group ({\it
solid line}: Lejeune et al. 1998 color transformations; {\it dashed
line}: Green et al. (1987) transformations), Padova group ({\it dotted
line}: Girardi et al. 2000), and Baraffe et al. (1998; {\it heavy
solid line}). The isochrones have been shifted using $(m-M)_V = 9.72$,
with an additional color shift (to account for reddening {\it and}
zero-point offsets in the color-$T_{\mbox{eff}}$ relations) to match the data
at $V \approx 14.5$. The color shifts were 0.03 in $V-I$ and 0.04 and
0.025 for the two Yale-Yonsei isochrones, 0.0 and 0.005 for Padova,
and 0.01 for Baraffe et al. (no $B-V$ models).
\label{isocomp}}
\end{figure*}

In the following discussion, we allow for small shifts in color to
accommodate the possibility of zero-point errors in the synthetic
colors or photometric calibration. The most robust comparison between
observations and theory appears to be using a color difference between
main sequence stars below the turnoff (where convective overshooting
plays a substantial role) and the warmest giants (at the base of the
giant branch). If we shift all of the isochrones so that they match
M67 stars at the absolute magnitude of the Sun (Fig. \ref{isocomp}),
we find that only the Y$^2$ isochrones using the LCB transformation
tables roughly reproduce the giant branch color: the Padova isochrones
predict giants that are too red by about 0.03 mag, and the Y$^2$
isochrones with GDK tables give giants that are too blue by about 0.06
mag. This is not too surprising because of the empirical
color-temperature relations used to correct the spectra used to
produce the LCB tables.  More surprising is the way that the Y$^2$-LCB
isochrones deviate more strongly from the observed main sequence than
the Y$^2$-GDK and Padova isochrones. In particular, LCB found that
uncorrected theoretical giant spectra deviated more strongly from
empirical relations (although mostly higher on the giant branch) than
did theoretical dwarf spectra. However, when the GDK tables are used,
the theoretical main sequence better matches the observations. In
Fig. 2 of Yi et al. (2001) comparing the GDK and LCB color
transformations, there is also a feature centered near $\log T_{\mbox{eff}} =
3.67$ that does not appear in any of the other color
transformations. We will discuss this further in \S \ref{msshape}
below.

\subsubsection{The Shape of the Turnoff\label{toshape}}

The largest effect on the shape of the turnoff region in M67 is the
algorithm used to model convective overshooting in the cores of
turnoff-mass stars. Overshooting can be described in terms of the distance
(in units of the pressure scale height $H_{P}$) a convective element
goes beyond the classical boundary defined by the Schwarzschild
criterion. In examining the \citet{mmj} dataset for M67,
\citet{dinescu} find that earlier models with overshoot of at most
$0.1 H_{P}$ seem to reproduce the observations best although they did
not discuss the details of their comparison.

\begin{figure*}
\includegraphics[width=144mm]{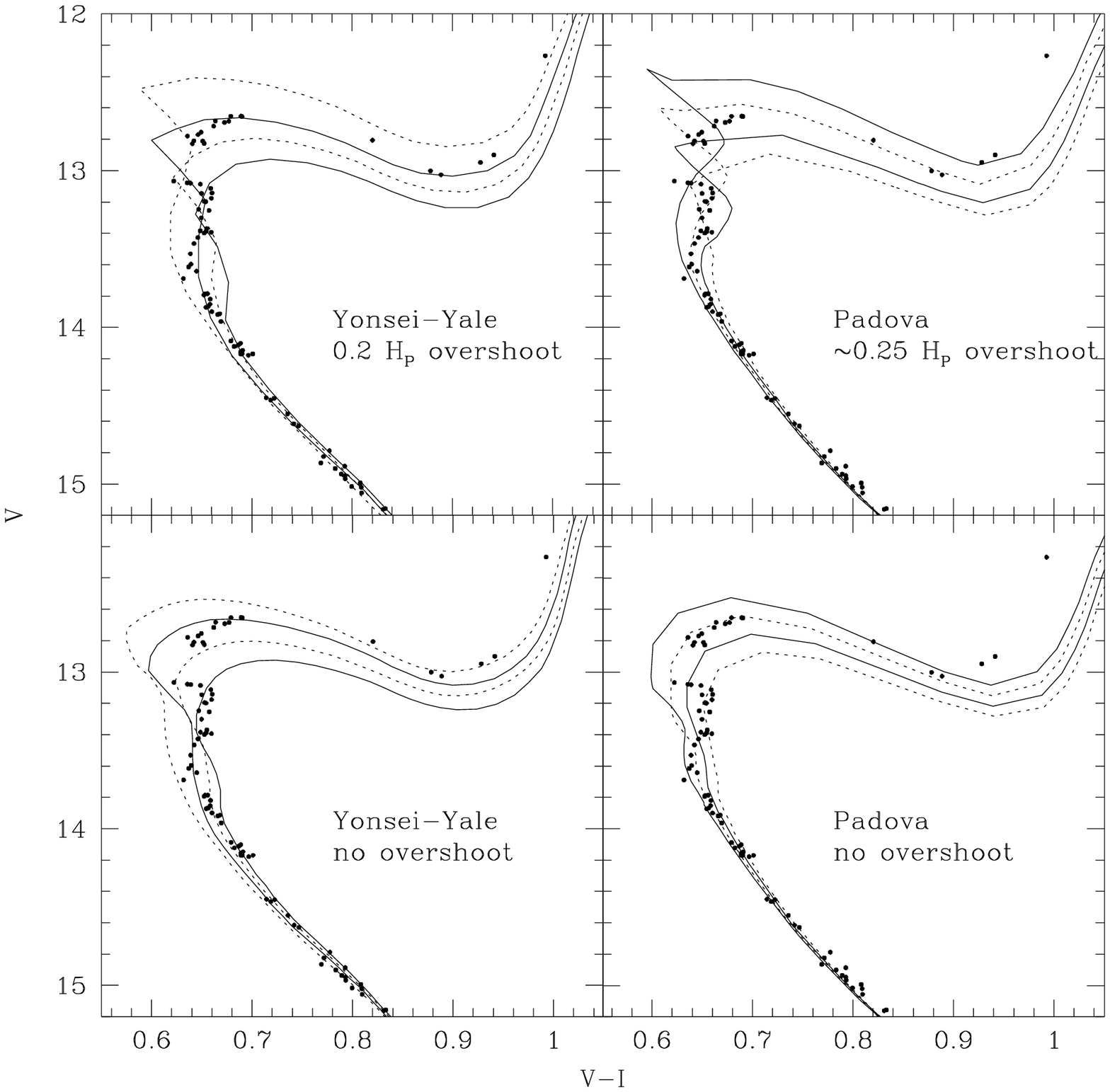}
\caption{Comparisons between our M67 observations and isochrones from
the Yonsei-Yale and Padova groups with and without convective core
overshooting. From brightest to faintest on the subgiant branch, the
isochrones are for ages 3.5, 4, 4.5 and 5 Gyr.
\label{tocomp}}
\end{figure*}

The Padova group isochrones and version 2 of the Y$^{2}$ isochrones
use similar algorithms to describe the overshooting: no overshooting
for low masses ($M < 1.0 \msun$ for Padova), constant overshoot for
higher mass stars ($0.2 H_{P}$ for Y$^2$, and approximately $0.25
H_{P}$ for $M \ge 1.5 \msun$ for Padova). Between these two extremes,
the amount of overshoot is ramped linearly. The overshooting parameter
in the Girardi et al. models ($\Lambda_{c}$) is described in Bressan,
Chiosi, \& Bertelli (1991; labelled there as $\lambda$), and describes
overshooting convective elements moving from {\it inside} the
convective zone, across the classical boundary, and into a radiative
region. This is not easily converted to the more typically quoted
extent of overshooting beyond the classical edge of the convection
zone because it involves the integration of a differential equation
for the velocity of a convective element. However, because the
acceleration of a convective element becomes negative at the classical
convection zone boundary, the overshoot beyond the boundary must be
less than one-half the mixing length (Maeder 1975). In the
\citet{girardi} models, $\Lambda_{c}$ varies as $\Lambda_{c} = (M /
M_{\odot}) - 1.0$ between 1.0 and $1.5 \msun$. Stars near the turnoff
mass in M67 have $M \approx $1.25$ \msun$, so that the amount of
overshoot beyond the zone boundary is less than about $0.13
H_{P}$. \citet{girardi} provide isochrones calculated without
convective overshooting, and version 1 of the Y$^2$ isochrones did not
use overshooting for ages similar to that of M67. In
Fig. \ref{tocomp}, we compare these four sets of isochrones to our
observed data for M67. Note that the amount of overshooting given in
the figure is for the higher mass range --- for stars at the turnoff,
the overshooting will be less.

The age of M67 makes it a particularly tricky test of the core
convection because M67 is at a stage in which the morphology of the
turnoff star evolutionary tracks changes rapidly as the extent of the
convective core decreases with decreasing mass. In addition, the
pressure scale height $H_{P} \longrightarrow \infty$ as $r
\longrightarrow 0$, so that the use of an algorithm forcing
overshooting of a constant fraction of $H_{P}$ is likely to break down
in this regime. \citet{rox} and \citet{woo2} have discussed ways of
constraining the amount of overshooting for stars with small
convective zones. In the majority of well-studied
clusters this is not an important issue, since an age more than a Gyr
different than M67's age is sufficient to make the typical overshoot
treatment adequate or to eliminate core convection from cluster
stars (e.g. Aparicio et al. 1990). In M67, this issue appears to be critical.

The morphology of the turnoff is the most easily used indicator of the
amount of overshooting in the current data. The major changes in
morphology from the overshooting are a more noticeable ``hook'' to the
red then blue prior to the SGB, and a change from a more vertical
(luminosity change) isochrone to a more horizontal (temperature
change) isochrone at the beginning of the SGB (the global $T_{\mbox{eff}}$
maximum). Woo et al. (2003) also discuss the color difference between the
turnoff and red giant branch as an indicator of the amount of
convective core overshoot. However, based on the results of \S
\ref{msrgb}, current uncertainties in color transformation render this
method somewhat questionable at present, particularly because it
relies on the {\it relative} accuracy of dwarf and giant atmosphere
models. 

The theoretical interpretation of the features is that the reddest
extent of the ``hook'' corresponds to the beginning of a rapid phase
of shrinkage of the extent of convection in the core of the star. At
the same time the core hydrogen content is quickly approaching
exhaustion, which occurs at the bluest point on the
isochrone. Following this, a hydrogen fusion shell source is
established around the core. Overshooting significantly affects
evolutionary time-scales, most notably lengthening the star's
main sequence life. It also affects the length of the beginning of the
subgiant phase by allowing more of the hydrogen {\it near} the centre
to be consumed, which requires a more substantial and rapid adjustment
of structure when the core hydrogen is exhausted. Thus, the
distribution of stars in the vicinity of the turnoff on a CMD carries
information about the amount of overshooting, although this can be
difficult to extract in the case of a relatively star-poor cluster
like M67.

There are other difficulties in interpreting M67's CMD, however.  In
terms of the observational morphology of the turnoff, we are faced
with a ``hook'' that appears to require some overshooting, but with an
extent less than seen in either of the theoretical isochrones with
overshooting. The redward-pointing portion of the hook is
well-populated and clearly at odds with all of the theoretical
isochrones without overshooting. If the four stars we identified in \S
\ref{lto} are truly single stars that are part of the
blueward-pointing portion of the ``hook'', this also can only be
explained with small amounts of overshooting. For example, fig. 5 of
Woo et al. (2003) presents synthetic CMDs incorporating different
amounts of overshooting for a somewhat younger cluster. Only the
models with overshooting of $0.1 H_{P}$ have a significant number of
single stars on the red half of the blueward-pointing part of the hook and
on the bluest parts of the SGB. In addition, higher amounts of
overshooting cause the red portion of the hook to overlap the
cluster's blend sequence, something which is not seen in
M67. However, the bluest portion of the subgiant branch appears to
most closely resemble the no-overshoot models, particularly those
of the Padova group.

\begin{figure}
\includegraphics[width=84mm]{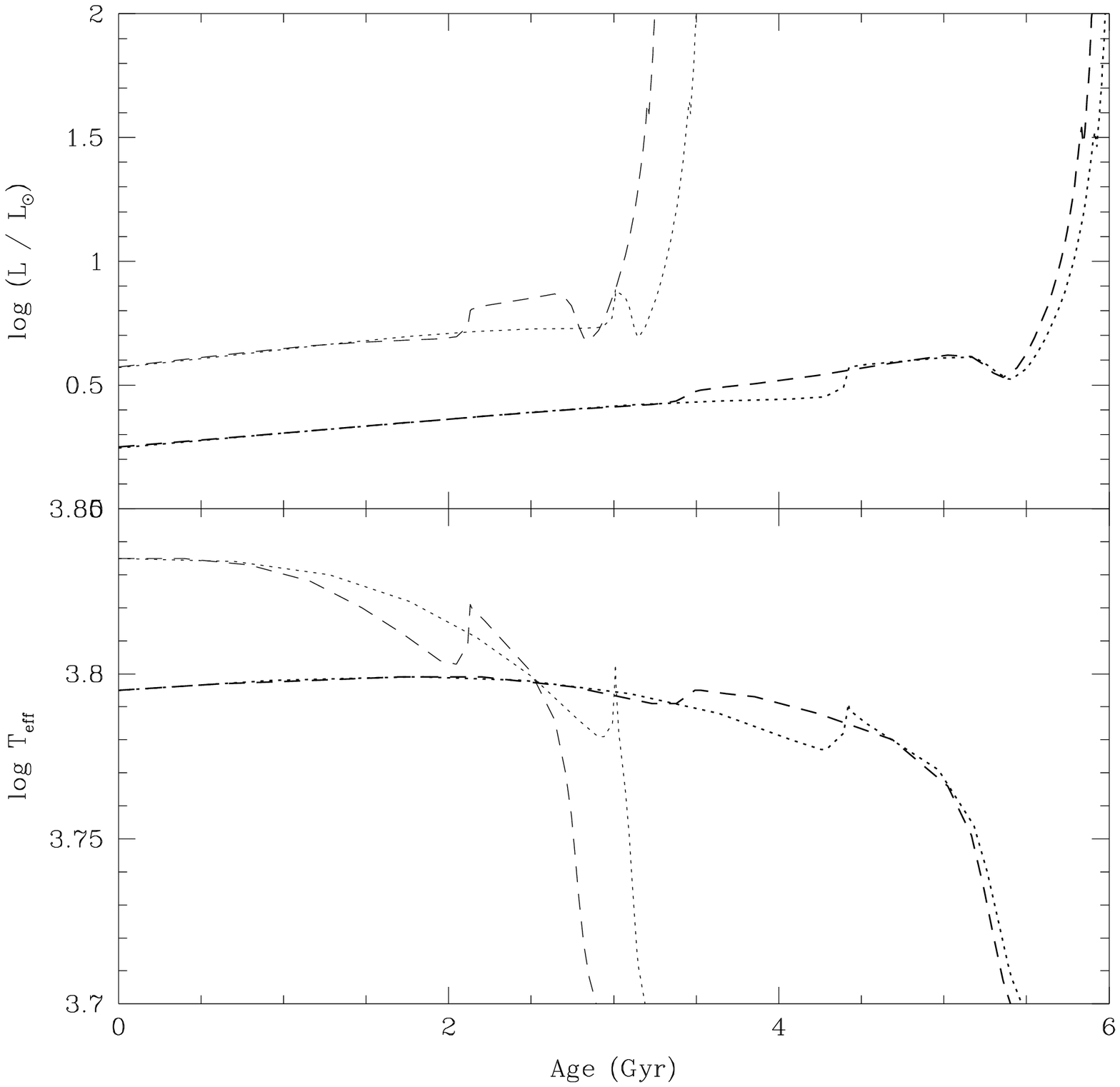}
\caption{The runs of $\log (L / L_{\odot})$ and $\log T_{\mbox{eff}}$ versus
age for $1.2 \msun$ ({\it heavy lines}) and $1.4 \msun$ ({\it thin
lines}) evolutionary tracks from the Padova group with ({\it dotted
lines}) and without ({\it dashed lines}) convective core overshooting.
\label{totheory}}
\end{figure}

The manner of our star selection makes it impractical to do a detailed
analysis of the numbers of stars in the CMD to examine issues of
evolutionary time-scales. However, the mere presence of the gap is
interesting.  Typically turnoff gaps seen in clusters like M67 have
been associated with the shrinkage of the convective core prior to
hydrogen exhaustion. However, models indicate that one of the most
noticeable differences when overshooting is introduced is the drastic
decrease in the amount of time spent on the bluest parts of the SGB
immediately after hydrogen exhaustion (see Fig. \ref{totheory}). If
the four stars are indeed on the blueward-pointing portion of the
hook, they give the impression that the evolutionary tracks should
evolve more in luminosity than in surface temperature at the beginning
of the SGB, and that the models should be evolving more quickly at the
beginning of the SGB to explain the gap.  This is crudely consistent
with presence of some overshooting, although the exact morphology of
the isochrone is still puzzling. Helium diffusion could play an
interesting role for M67 turnoff stars by helping to bring additional
hydrogen fuel toward the core. The models of the Padova group do not
include helium diffusion, while those of the Y$^2$ group do.

Our best interpretation of the turnoff morphology is that a fairly
small convective core including some amount of convective overshoot is
necessary, but that the extent of the mixed core at hydrogen is
small. Because the evolutionary time-scale for the stars increases
dramatically once the hydrogen fusion regions reach hydrogen rich gas
outside of the regions earlier occupied by a convective core, the
presence of a gap puts a lower limit on the extent of the core. If the
convective core in a turnoff star is never particularly large, then
after core hydrogen exhaustion the situation is similar to that of a
lower mass star. If there is not a large region of hydrogen-exhausted
gas near the core, then the star does not need to adjust its structure
much to find a new, stable configuration. In this sense, the gap in an
old cluster like M67 can be said to be the last remnant of the
Hertzsprung gap seen in much younger open clusters: when core hydrogen is
exhausted in turnoff stars in those clusters, a greater degree of
structural readjustment is necessary.

Our interpretation of these features does depend on a small number of
stars, some of which could be blue stragglers or unresolved
binaries. If all or most of them are single stars though (as we
believe), the morphology of the turnoff deserves a more detailed look
because it can tightly constrain the behavior of gas in small
convective cores. 

The exact amount of overshoot has only a small effect on the age inferred from
M67's turnoff. However, 4 Gyr isochrones reproduce most aspects of the
upper main sequence and subgiant branch well, so we regard this as the
preferred age, and estimate that the uncertainty is less than about
0.5 Gyr. This is in agreement with the value derived by
\citet{dinescu}, although they used an earlier set of models and
different photometry.

\subsubsection{The Shape of the Main Sequence\label{msshape}}

Using $V-I$ color, the Y$^2$-GDK and Padova isochrones agree with each
other and satisfactorily match the shape of the main sequence nearly
to $V = 16$. The Y$^2$-LCB isochrone diverges rather substantially
from the M67 SSS by $V = 15.5$. In all three cases though, the main
sequence models become far too blue. We find that the
solar-metallicity models of BCAH provide a much better fit to the
shape of the main sequence down to the limit of our calibration at $V
=17.8$. In $B-V$ color, the Padova and Y$^2$-GDK isochrones still
agree, but the Y$^2$-LCB isochrones is a much closer fit to the actual
data to $V \approx 16.75$ (see Fig. \ref{isocomp}). This is a clear indication
that the color-$T_{\mbox{eff}}$ relations are responsible for much of the
disagreement when it comes to fitting the main sequence.

One of the reasons the main
sequence color calibration is so difficult is the lack of a good,
independently-determined $T_{\mbox{eff}}$ scale for faint dwarfs. Houdashelt,
Bell, \& Sweigart (2000) examined data for cool (4000 K $\le T_{\mbox{eff}}
\le$ 6500 K) field stars and found indications that the
color-$T_{\mbox{eff}}$ relations for dwarfs and giants diverge for $T_{\mbox{eff}} <
5000$ K in $(V-R)_C$ and $(V-I)_C$ colors. According to their data,
the dwarf stars become redder than the giants at a given $T_{\mbox{eff}}$,
which is in the correct direction to explain the discrepancies between
models and observations of M67. This could be the result of
gravity-sensitive molecular features like TiO and CN that fall in $V$,
$R$, and $I$ bands (Houdashelt et al. 2000). The empirical
color-$T_{\mbox{eff}}$ relations used to correct the models of LCB (Bessell
1979, 1995) are systematically bluer than the Houdashelt et al. field
dwarf values (see their fig. 12), and Bessell (1979) actually used
field giant data to set the relation for cool dwarfs, assuming the two
are equivalent. But as Houdashelt et al. state, systematic errors in
$T_{\mbox{eff}}$ measurements for the coolest dwarfs could also be
responsible for the differences between field dwarf and field giant
color-$T_{\mbox{eff}}$ relations.

According to empirical relations, $T_{\mbox{eff}} = 5000 K$ corresponds to
$(V-I)_0 = 0.93$. In our M67 data, this roughly corresponds to the
color at which the Padova and Y$^2$-GDK isochrones begin diverging
from the bluest stars on the M67 main sequence. The Y$^2$-LCB models
diverge closer to $(V-I) = 0.85$, which appears to be due to a lack of
empirical data to calibrate the $(V-I)-T_{\mbox{eff}}$ relation (P. Westera,
private communication).

The primary differences in input physics between the best-fitting
models of BCAH and those of the Y$^2$ and Padova groups are in the
equation of state and in the surface boundary conditions. BCAH use
recent non-grey stellar atmosphere models as boundary conditions for
their stellar models and the equation of state of Saumon, Chabrier, \&
Van Horn (1995; hereafter SCVH). The Padova isochrones make use of the
MHD equation of state \citep{mhd} and opacities from Alexander \&
Ferguson (1994) for surface layers. The Y$^2$ isochrones also use
\citet{alex} opacities at the stellar surface. For the equation of
state they use the OPAL \citep{opal} tables at temperatures down to
$\log T = 3.7$, and then revert to a Saha equation of state with
Debye-H\"{u}ckel correction for Coulomb interactions.

SCVH have shown that significant differences exist between their
equation of state and others, and that the MHD or OPAL equations of
state are the most accurate for stars of solar mass or above while the
SCVH equation of state is preferred for low mass stars due to its
inclusion of non-ideal effects and molecule formation. However,
\citet{cb} indicate that comparisons of models calculated using the
SCVH and MHD equations of state show agreement to better than 1.3\% in
$T_{\mbox{eff}}$ and $L$ for stars with $M \ga 0.4 \msun$.  The Y$^2$ models
used a simpler equation of state at low temperatures ($\log T <
3.7$). The Y$^2$ models reach this threshold in the surface layers for
stars with mass near $0.81 \msun$ ($M_{V} \approx 6.3$), which appears
to be at a point on the main sequence that is too faint to explain the
deviations.  So, our feeling is that current equation of state
differences are not likely to be of high importance in the range of
star masses we have observed.

At low $T_{\mbox{eff}}$, studies (e.g. Chabrier \& Baraffe 1997) have
forcefully shown the importance of the surface boundary condition in
determining the luminosity and spectral type of the star. For fully
convective stars, the temperature gradient is forced to be very close
to the adiabatic temperature gradient, so that the surface conditions
(where the adiabaticity breaks down) essentially set the entire
interior structure of the star. Unfortunately, stellar atmospheres at
low $T_{\mbox{eff}}$ are strongly non-grey (e.g. Chabrier \& Baraffe 1997),
so that to adequately determine the interior structure, detailed
stellar atmospheres have to be used as surface boundary
conditions. While the BCAH models diverge from deep photometry of open
clusters for $T_{\mbox{eff}} < 3700$ K \citep{cb}, this limit is fainter on
the main sequence than either the Padova or Y$^2$ have been tested. In
addition, BCAH use color transformation tables derived in a self-consistent
way from their model atmospheres while the Padova and Y$^2$ isochrones
use tables derived from atmospheres that are not applied as surface
boundary conditions.  These are the primary reasons behind the different
CMD positions of the BCAH main sequence and those of the Padova and
Y$^2$ groups.

Differences between the isochrones due to the assumed equation of
state, color-temperature transformations, and surface boundary
conditions get severe the fainter on the main sequence the models go
(e.g. von Hippel et al. 2002). As yet, the Y$^2$ and Padova isochrones
have not been rigorously tested in this regime (see Yi et al. 2001),
so it is not entirely fair to expect agreement as yet. However, when
the models are shifted to match up with the observed data at the
magnitude level of the Sun (thereby differentially removing small
effects of the differences in age and composition, and the larger
effects of differences in color zero points) all of the models except
those of BCAH are clearly too blue. \citet{vonH} found a similar
discrepancy extending to fainter levels for the cluster M35. 
However, M67 has a composition that is closer to that of the Sun, and so 
{\it should} closely match the solar neighborhood stars used to calibrate 
color-$T_{\mbox{eff}}$ relations.

\section{Conclusions}

We have used the extensive database of observations we have gathered
in our variability studies to produce a high-precision CMD for the old
solar-abundance open cluster M67. The age of M67 has created turnoff
stars that are showing the effects of overshooting in their small
convective cores. M67 thereby provides a severe test of algorithms used
to model stars with masses $1.1 \la M / \msun \la 1.3$.

If M67 data is calibrated to fainter magnitudes in the future, it will
be possible to more strongly test the physics involved in the
computation of the stellar atmospheres of main sequence
dwarfs. Because of the solar abundance of the cluster, M67 can be
extremely valuable in constraining color transformations for main
sequence stars of solar abundance. Because solar abundance stars often
provide the basis for determining the transformations at lower
abundances, the data for M67 has wide application.

The accurate determination of the lower main sequence fiducial line would
also open the possibility of predicting the properties of both stars
in unresolved binaries having mass ratios with $q \la 0.7$. In this
range, a binary system can be uniquely deconvolved into its components
under the assumption that there are just two non-interacting
stars. Our photometry does not go deep enough down the main sequence
to allow us to make this deconvolution without additional data. With
just the addition of a reasonably good theoretical main sequence line,
it would be possible to determine a photometric mass ratio
distribution for a fairly large sample of stars in this
cluster. Because M67 appears to be a cluster that is quite relaxed
dynamically, this would give us the opportunity of directly seeing how the
cluster dynamics have affected the binary stars.

\section*{Acknowledgments}

E.L.S. would like to thank the director of Mount Laguna Observatory
(P. Etzel) for generous allocations of telescope time that made this
study possible, M. Shetrone for many useful conversations, M. van den
Berg for very useful conversations and for thorough comments on the
manuscript, P. Westera for insight into $T_{\mbox{eff}}$-color
relations, and the anonymous referee for valuable comments about
mechanisms broadening a single star sequence that resulted in \S
\ref{broad}. This research has made use of the SIMBAD database
operated at CDS, Strasbourg, France, and was supported in part by the
National Science Foundation under Grant No. AST-0098696.

\end{document}